%% file: main.tex
\documentclass[12pt]{iopart}
\usepackage[utf8]{inputenc}
\usepackage{pgfplots}
\usepackage{color, colortbl}
\usepackage{import}
\expandafter\let\csname equation*\endcsname\relax
\expandafter\let\csname endequation*\endcsname\relax
\usepackage{amsmath}
\usepackage{amssymb}
\usepackage{tabularray}
\usepackage{mwe}
\usepackage{graphbox}
\usepackage{harvard}
\usepackage{url}
\usepackage{hyperref}
\usepackage{xr-hyper}
\pgfplotsset{compat=1.18}
\citationmode{abbr}
\definecolor{light-gray}{gray}{0.8}
\definecolor{table-gray}{gray}{0.9}

\newcommand{\brm}[1]{\boldsymbol{\mathrm{#1}}}

\begin{document}

\title[Anatomically Informed GREIT Reconstruction in EIT Lung Monitoring ]{Anatomically Informed GREIT Reconstruction: Improving EIT Imaging for Lung Monitoring }

\author{Maximilian Ludwig$^{1}$ , Carolin M. Eichinger$^{1}$ , Armin Sablewski$^{2}$, Inéz Frerichs$^{2}$, Tobias Becher$^{2}$, Wolfgang A. Wall$^{1}$ }

\address{$^{1}$ Institute for Computational Mechanics, Technical University of Munich, Garching b. München, Germany}
\address{$^{2}$ Department of Anesthesiology and Intensive Care Medicine, University Medical Center Schleswig-Holstein, Kiel, Germany}
\ead{maximilian.ludwig@tum.de}
\vspace{10pt}

\begin{abstract}
\textit{Objective:} Time-difference electrical impedance tomography (EIT) is gaining widespread use for bedside lung monitoring in intensive care patients suffering from lung-related diseases. It involves collecting voltage measurements from electrodes placed on the patient’s thorax, which are then used to reconstruct impedance images. This study investigates how incorporating anatomical information from CT data into the widely used GREIT reconstruction algorithm affects EIT images and improves their interpretability. \textit{Approach:} Based on clinically motivated lung state scenarios, we simulated EIT measurements to assess how the GREIT parameters influence the result of EIT image reconstruction, particularly with respect to noise performance and image accuracy. We introduce quality measures that allow us to perform a quantitative assessment of reconstruction quality. We incorporate the anatomical features of a patient from CT data by customizing the background conductivity and the distribution of GREIT training targets. \textit{Main results:} Our analysis confirmed that unphysiological background conductivity assumptions can lead to misleading EIT images, whereas physiological values, although more accurate, come with higher noise sensitivity. By increasing the number of GREIT training targets inside the lung and adapting the respective weighting radius, we significantly improved the anatomical accuracy of the EIT images. When applied to clinical EIT data from a representative ARDS patient, these adjustments in the reconstruction setup substantially enhanced the interpretability of the resulting EIT images. \textit{Significance:} Incorporating CT-based anatomical data in the GREIT reconstruction significantly enhances the clinical applicability of EIT in lung monitoring. The improved interpretability of EIT images facilitates better-informed clinical decisions and the individualized adjustment of ventilation strategies for critically ill patients.
\end{abstract}
\vspace{2pc}
\noindent{\it Keywords}: electrical impedance tomography, image reconstruction, lung monitoring, GREIT, anatomical accuracy

\section{Introduction}
The application of electrical impedance tomography (EIT) in functional lung monitoring leverages the dependency between electrical impedance and local aeration of lung tissue to gain insight into the regional distribution of lung ventilation. In contrast to common pulmonary monitoring techniques, where only global quantities are measured, insight into the regional ventilation distribution obtained by EIT holds promises towards application of more protective ventilation strategies to prevent ventilator-induced lung injury. The radiation-free nature of EIT allows for bedside monitoring of critically ill patients, effectively bridging the gap between classical lung monitoring and CT imaging with its high spatial resolution. This enables more individualized ventilatory patient treatment. Particularly for patients with heterogeneous lung states, such as those as occur in acute respiratory distress syndrome (ARDS), the regional information is crucial to recruit collapsed lung regions while simultaneously keeping overdistension minimal \cite{Costa2009,Jonkman2023} and to improve ventilation-perfusion matching \cite{Yang2025}.  In the common clinical use case of functional lung monitoring, EIT is mainly applied as time-difference EIT due to its higher robustness against model inaccuracies and measurement noise \cite{Kobylianskii2016,Frerichs2017,Scaramuzzo2024}. In this context, EIT image signals need to be interpreted as the change in aeration with respect to some reference state, which is commonly set to the end-expiratory level. \\
A recent application of EIT involves its use as a tool for the validation of predictive, physics-based computational lung models, as demonstrated by \citeasnoun{Roth2017} and \citeasnoun{Rixner2024}. This application benefits from the non-invasive and bedside-compatible nature of EIT for a validation of the personalized computational lung models in a clinical setting. Accordingly, the spatial accuracy of the reconstructed EIT images is essential for validating the spatial detail of the model predictions.\\ 
Mathematically speaking, calculating the impedance change image from the electrode voltage differences is an ill-posed inverse problem. Therefore, reconstruction algorithms require regularization techniques to obtain plausible impedance distributions. The most commonly used algorithm for EIT image reconstruction is the Graz consensus Reconstruction algorithm for EIT (GREIT) \cite{Adler2009}. In several studies comparing reconstruction algorithms, GREIT has shown a very good - and often the best - performance regarding common reconstruction accuracy measures \cite{Adler2009,Thurk2017,Zhao2014,Grychtol2014}. \\ 
However, some uncertainties in parameterization and commonly used simplifications influence the EIT images obtained with GREIT. One of these is the background conductivity assumption underlying the generation of GREIT training data. The training data result from evaluations of the forward model where non-conductive targets are distributed in a domain with a certain background conductivity distribution and compared to a reference measurement without the placement of non-conductive targets. The background conductivity can be interpreted as the linearization point for determining the Jacobian and, therefore, has a crucial influence on the reconstruction. \citeasnoun{Grychtol2013} discovered that the commonly applied uniform background conductivity distribution produces misleading EIT images. Hence, they suggest employing a more physiological distribution of the background conductivity to achieve higher accuracy and reliability of EIT images. Another common simplification in the reconstruction of EIT images is the application of simplified thorax geometries, such as circular or ellipsoid shapes. Particularly in clinical EIT monitoring devices, the common choice for the model shape is the usage of population-averaged thorax geometries. However, this mismatch between real and modeled geometry can be considered a modeling error and thus produce misleading images \cite{Grychtol2012}. Both effects - the influence of the background conductivity as well as that of the model geometry - were investigated by \citeasnoun{Thurk2017}, comparing EIT reconstructions to 4D-CT, which can be considered the gold standard for obtaining information on temporal and regional ventilation of the lung. They have shown that EIT image quality benefits significantly when anatomical information is considered during reconstruction. In a later study, \citeasnoun{Thurk2019} validated their reconstruction setup using a functional and physiological validation framework, which was first established by \citeasnoun{Grychtol2014}. \\ 
The present study aims to improve the reliability of reconstructed EIT images and enhance anatomical accuracy by incorporating anatomical features into the reconstruction setup. We achieve this by leveraging geometrical information from single X-ray CT scans which are part of the clinical routine when treating lung insufficiency. One parameter in the GREIT algorithm that is already known to influence the anatomical accuracy of EIT images is the background conductivity distribution. We further investigate this influence on the accuracy and assess the impact on noise performance using artificial ventilation scenarios that are motivated by realistic lung states in ARDS and serve as a well-defined ground truth. The second option to consider anatomical features in the reconstruction approach is to adapt the distribution of training targets in GREIT from the commonly uniform distribution to a customized, anatomically informed distribution. This is expected to increase the accuracy of the EIT images in the region of interest, which in our case is the lung. Again, we assess the influence of training target related parameters employing a set of artificial ventilation scenarios, where we have a well-defined ground truth. With the anatomically informed reconstruction approach presented in this study, we aim to improve the interpretability of clinical EIT images enabling more individualized ventilatory treatment of critically ill patients in the intensive care unit.
\section{Methods}
To detail our methodology for finding a suitable parameterization of GREIT, we first recall the fundamentals of EIT forward and inverse modeling, see Sections \ref{sec:fwd_model} and \ref{sec:GREIT_fundamental}, respectively. We then adapt the commonly used GREIT approach by varying the distribution of the background conductivity (Section \ref{sec:background_conductivity}) and  introducing a region-dependent target distribution (Section \ref{sec:custom_target_distr}). In Section \ref{sec:scenario_setup_method}, we propose an artificial simulation setup to evaluate the EIT forward problem on a simplified but physiologically motivated thorax model. After distorting the resulting voltage data with uncorrelated Gaussian noise, we assess the reconstruction quality in terms of accuracy and robustness with respect to noise (Section \ref{sec:quality_measures}).
\subsection{Forward Model}\label{sec:fwd_model}
The EIT forward model predicts the electric potential field $u$ given the distribution of the electrical conductivity $\sigma$. This assumes that the imaginary component of the complex-valued electrical admittivity can be neglected for the frequency range of pulmonary EIT \cite{Simini2018}. Considering the absence of interior current sources, the electrical current density $\brm{J} = -\sigma\nabla u$, given by Ohm's law, is divergence-free \cite{Lionheart2005}. This results in the elliptic partial differential equation  
\begin{equation}\label{eq:laplace}
    \nabla \cdot\sigma \; \nabla u = 0
\end{equation}
governing the electric potential $u$ within the domain $\Omega$. Together with the Complete Electrode Model \cite{Cheng1989,Somersalo1992}, which characterizes the boundary conditions on the electrode surfaces, the electric potential is sufficiently described.\\
In the present work, the EIDORS toolbox\footnote[6]{\url{https://eidors3d.sourceforge.net/}} version 3.10 is used to discretize and solve the governing equations above using the finite element method in the three-dimensional domain. This step can be abstracted and written in terms of a forward operator $F$. Instead of using the electric potential field $u$ in the entire domain, we formulate the forward operator in terms of the discrete electrode voltages that can be written in vector-form as $\brm{v} \in \mathbb{R}^{n_M}$. $n_M$ refers to the number of voltage measurements and depends on the number of electrodes and their connection pattern for stimulation and measurement. The conductivity distribution is discretized by $n_\mathrm{ele, 3D}$ three-dimensional finite elements resulting in the vector-form for the discrete conductivity $\brm{\sigma} \in \mathbb{R}^{n_\mathrm{ele, 3D}}$. We can thus write the forward model operator as $F: \mathbb{R}^{n_\mathrm{ele, 3D}} \mapsto \mathbb{R}^{n_M}, \;\brm{\sigma} \mapsto \brm{v}$. For a detailed derivation of the formulation of the forward model $F$, the reader is referred to the corresponding literature on EIDORS \cite{Polydorides2002,Adler2006}.
\subsection{GREIT Formulation}\label{sec:GREIT_fundamental}
The classical EIT reconstruction problem goes in the inverse direction compared to the forward model operator $F$. To reconstruct the electrical conductivity distribution given the electrode voltage measurement we employ the commonly used GREIT algorithm based on a generalization of the Tikhonov regularization \cite{Adler2009}.  In fact, we consider the voltage difference $\brm{y} = \brm{v} - \brm{v}_0$ with respect to the reference state $(\brm{v}_0, \,\brm{\sigma}_0)$, to reduce the effects of model inaccuracies and data noise. Thus, the reconstructed image need to be interpreted as the conductivity change rather than the absolute conductivity. \\
In EIT image reconstruction, the inverse mapping from the voltage difference to the reconstructed image of the conductivity change $\brm{\hat{x}} \in \mathbb{R}^{n_\mathrm{ele, 2D}}$ is approximated by the linear formulation 
\begin{equation}\label{eq:linear_reconstruction}
\brm{\hat{x}} = \brm{R}\,\brm{y}, 
\end{equation}
where $\brm{R} \in \mathbb{R}^{n_\mathrm{ele, 2D} \times n_M}$ represents a reconstruction matrix. Note, that we distinguish between the number of elements in the three-dimensional forward model $n_\mathrm{ele, 3D}$ and those in the reconstructed two-dimensional slice $n_\mathrm{ele, 2D}$, where typically $n_\mathrm{ele, 3D} > n_\mathrm{ele, 2D}$. For most clinical applications the images can not be reconstructed in three dimensions as the electrodes are typically positioned in a planar arrangement. As mentioned above, we use the GREIT algorithm to obtain the reconstruction matrix $\brm{R}$. To recall the definition, we use the formulation of \citeasnoun{Grychtol2016}. They describe $\brm{R}$ based on the definition of $N_t$ "training" conductivity distributions $\brm{t}_{i} \in \mathbb{R}^{n_\mathrm{ele, 3D}}$. Each of these training data sets corresponds to the placement of a single spherical non-conductive target inside the domain with a given background conductivity. For each of these training conductivity distributions one can evaluate the forward model resulting in a training measurement $\brm{\tilde{y}}_i$. Additionally one can formulate a desired image $\brm{\tilde{x}}_i\in \mathbb{R}^{n_\mathrm{ele, 2D}}$ for each training sample $\brm{t}_{i}$. Following \citeasnoun{Grychtol2016}, we model the desired images of sample $i$ by a sigmoid function of a certain radius $R_{w,i}$ which is commonly referred to as weighting radius. The reconstruction matrix in GREIT is then formulated in terms of these training data and seeks to minimize the error 
\begin{equation} \label{eq:greit_error_min}
    \epsilon^2(\brm{R}) = \underset{w}{\mathrm{E}} \left[ \left\| \brm{\tilde{x}}-\brm{R}\brm{\tilde{y}}\right\|^2 \right]
\end{equation}
where $\underset{w}{\mathrm{E}} [x] = \frac{1}{N_t} \sum^{N_t}_{i=1} w_i\,x_i$ represents the expectation over the set of $N_t$ targets weighted by $w$ \cite{Adler2009,Grychtol2016}. The minimization results in the following formulation for $\brm{R}$
\begin{equation} \label{eq:greit_reconstruction_matrix}
    \brm{R} = \underset{w}{\mathrm{E}} \left[ \brm{\tilde{x}}\brm{\tilde{y}}^{\mathrm{T}} \right] \left(\underset{w}{\mathrm{E}} \left[ \brm{\tilde{y}}\brm{\tilde{y}}^{\mathrm{T}} \right] \right)^{-1},
\end{equation}
with
\begin{align}
    &\underset{w}{\mathrm{E}}\left[ \brm{\tilde{y}}\brm{\tilde{y}}^{\mathrm{T}} \right] = \frac{1}{N_t} \sum\limits_{i=1}^{N_t} w_i\,\brm{\tilde{y}}_i\brm{\tilde{y}}_i^\mathrm{T} + \frac{1}{N_n} \sum\limits_{i=1}^{N_n} \brm{n}_i\brm{n}_i^{\mathrm{T}}, \label{eq:greit_expect_yy}\\
    &\underset{w}{\mathrm{E}}\left[ \brm{\tilde{x}}\brm{\tilde{y}}^{\mathrm{T}} \right] = \frac{1}{N_t} \sum\limits_{i=1}^{N_t} w_i\,\brm{\tilde{x}}_i\brm{\tilde{y}}_i^\mathrm{T}. \label{eq:greit_expect_xy}
\end{align}
The term $\frac{1}{N_n}\sum_{i=1}^{N_n} \brm{n}_i\brm{n}_i^{\mathrm{T}} = \brm{\Sigma}_n$ models the measurement noise based on $N_n$ measurement sets~$\brm{n}_i$ consisting of pure noise. It is common practice to use one single uncorrelated noise measurement set so that the noise covariance becomes a scaled identity matrix $\brm{\Sigma}_n = \lambda^2\, \brm{I}$. $\lambda$ represents the reconstruction hyperparameter and manipulates the degree of regularization of the reconstruction algorithm. The optimal value of the hyperparameter $\lambda$ depends heavily on the specific characteristics of the problem at hand, which is why in GREIT the user typically does not specify $\lambda$ explicitly, but rather prescribes a certain noise amplification behavior referred to as noise figure $\mathrm{NF}$ \cite{Adler2009}. $\mathrm{NF}$ is determined by comparing the reconstructed image of a single spherical non-conductive target in the center of the domain to the image of a measurement of pure noise, see \citeasnoun{Adler2009}.\\
In the following, we consider the weight in equations \eqref{eq:greit_expect_yy} and \eqref{eq:greit_expect_xy} to be uniform $w_i=1$ as a heterogeneous distribution of the weight is not implemented in GREIT so far. However, by distributing the training targets $\brm{t}_i$ heterogeneously over the domain we can manipulate the influence of a certain region on the reconstruction \cite{Grychtol2016}. \\
\subsection{Incorporating Anatomical Information in the Reconstruction}
In many clinical scenarios, such as the intensive care treatment of ARDS patients, geometrical data on thorax and lung topology can be extracted from X-ray CT scans, which are performed in clinical routine. Leveraging this data, we aim to enhance reconstruction methods in terms of noise performance and reliability.
\subsubsection{Background Conductivity.} \label{sec:background_conductivity}
\begin{figure}
    \centering
    \fontsize{12pt}{11pt}\selectfont
    \def\svgwidth{3.333in}
    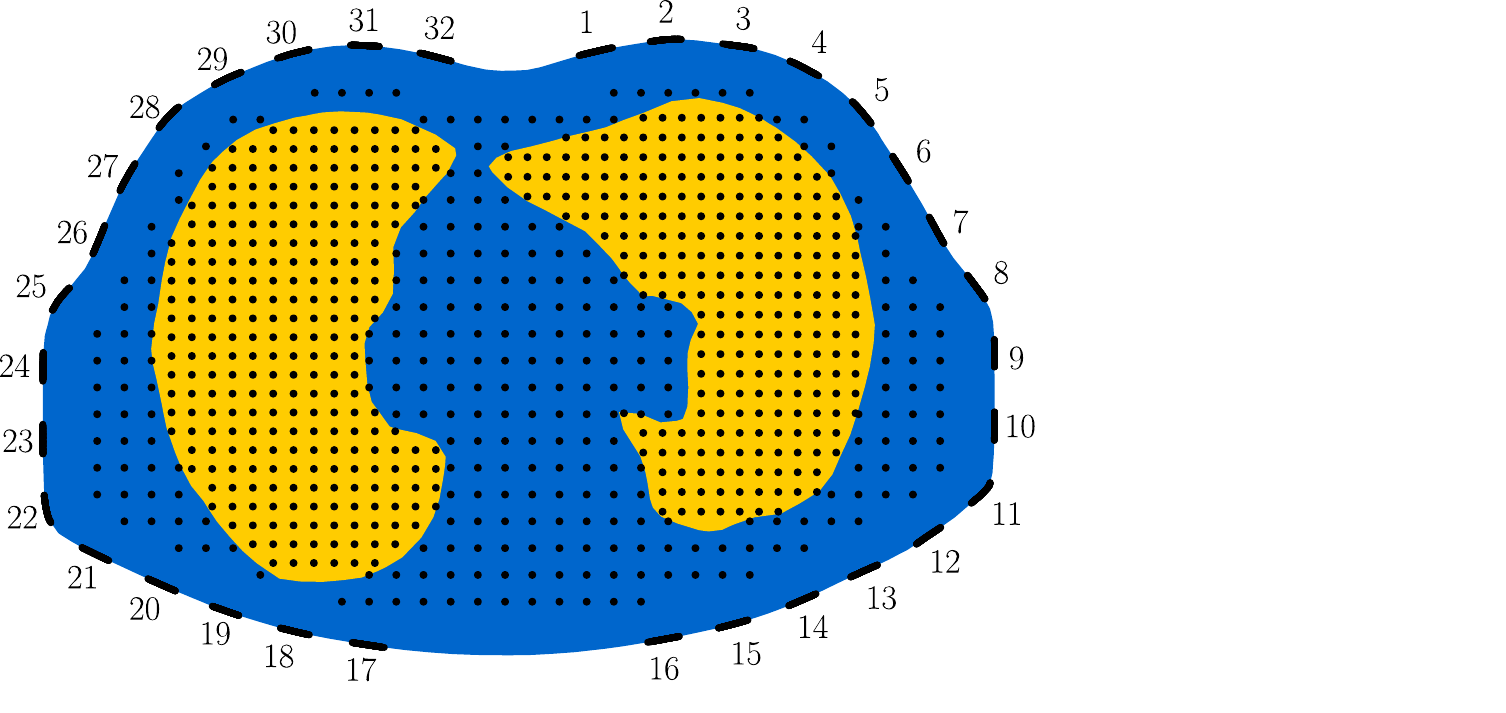
    \caption{Regionally homogeneous parameter setup to generate GREIT training samples. The background conductivity of the lungs $\sigma_{\mathrm{bkg, lungs}}$ and that of the thorax $\sigma_\mathrm{bkg, thorax}$ are illustrated by the colors yellow and blue, respectively. The black dots indicate the position of the non-conductive targets. The spatial densities of targets inside the lung and the thorax may differ. No targets are placed within a distance of $s_\mathrm{bound}$ from the thorax boundary.}
    \label{fig:backg_cond_target_distr}
\end{figure}

The linear reconstruction matrix $\brm{R}$ essentially depends on the forward model and, more precisely, on the Jacobian matrix $\brm{J}$, which is a linearization of the system around a reference state. In thoracic EIT, it is common practice to assume a homogeneous conductivity distribution across the whole domain as the reference for the linearization, although this is a clear deviation from the heterogeneous reference state in reality. \citeasnoun{Grychtol2013} suggest using a more realistic reference to determine the linearization point by taking the contours and the conductivity of the lungs and the heart into account. \\
In our work, we follow on from \citeasnoun{Grychtol2013} and aim to reduce the reference point mismatch by considering the lung topology and setting a regionally homogeneous conductivity for lungs, $\sigma_{\mathrm{bkg, lung}}$, and the thorax (i.e., the non-lung region), $\sigma_{\mathrm{bkg, thorax}}$. Although the regionally homogeneous assumption is not the most sophisticated way to parameterize the background conductivity, it allows quantitative parameter studies on the influence of the background conductivity assumption, as we will show in Section \ref{sec:param_study_unif}. \Fref{fig:backg_cond_target_distr} illustrates the regionally homogeneous background conductivities of the lungs and the thorax in yellow and blue, respectively. To describe the background conductivity in the following sections, we will mainly refer to the ratio
\begin{equation} \label{eq:background_conductivity_ratio}
    \gamma_\sigma = \frac{\sigma_{\mathrm{bkg,\,lung}}}{\sigma_{\mathrm{bkg,\,thorax}}}.
\end{equation}
\subsubsection{Custom Target Distribution.} \label{sec:custom_target_distr}
Another investigated adaptation of the common GREIT workflow is the application of a customized distribution of training targets. \citeasnoun{Grychtol2016} state that considering a non-uniform target distribution can achieve similar results as a heterogeneous weight distribution in Equation \eqref{eq:greit_error_min}. \\
By choosing the number of training targets per area, i.e., target density, depending on the anatomical region (higher within the lungs, lower outside the lungs), we aim to force the GREIT optimization to interpret changes in electrode voltages increasingly as the result of an impedance change in the lung region instead of changes in the rest of the thorax. As for the background conductivity distribution above, we use a regionally homogeneous spatial target density for the lungs $\rho_{\mathrm{lung}}$ and the thorax $\rho_{\mathrm{thorax}}$ which can be condensed in the ratio 
\begin{equation} \label{eq:target_density_ratio}
\gamma_\rho = \frac{\rho_{\mathrm{lung}}}{\rho_{\mathrm{thorax}}}.\\
\end{equation}
Due to the high sensitivity of the forward problem to conductivity changes close to the electrodes, training targets at the thorax boundary might cause artifacts in the impedance reconstructions. Therefore, we remove training targets closer to the thorax boundary than some distance threshold $s_{\mathrm{bound}}$. Figure \ref{fig:backg_cond_target_distr} illustrates the customized target distribution and the parameters defining it.\\
\subsection{Evaluation Procedure to Study Reconstruction Performance}\label{sec:scenario_setup_method}
\begin{figure}
    \centering
    \fontsize{10pt}{11pt}\selectfont
    \def\svgwidth{6.2in}
    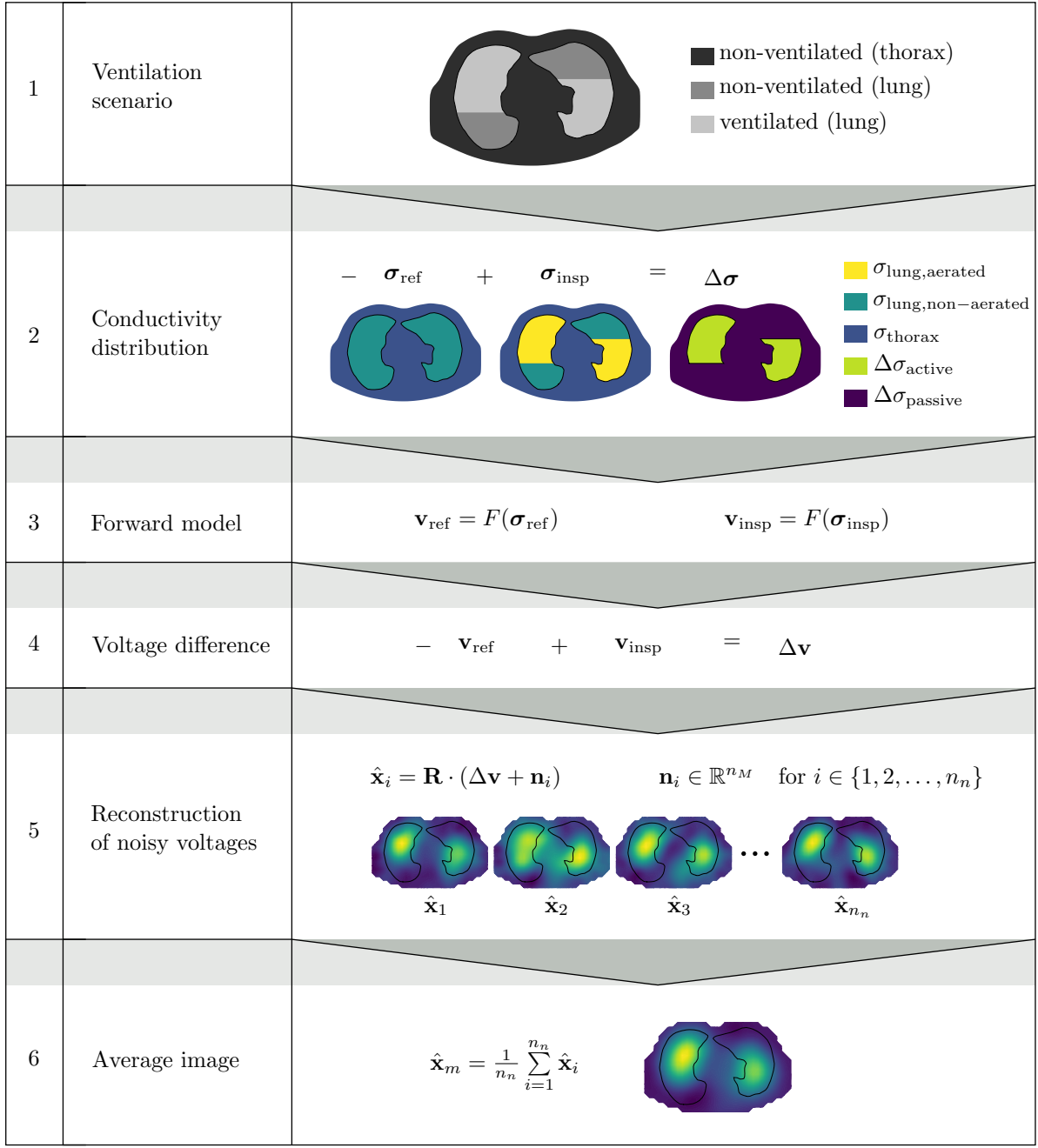
    \caption{Evaluation procedure for a given ventilation scenario. Samples from a modeled measurement noise $\brm{n}_i$ distort the simulated noise-free voltages. }
    \label{fig:scenario_methods}
\end{figure}

To investigate the influences of the different reconstruction approaches and parameterizations, we consider simulated electrode voltage measurements, which we impair with additive measurement noise. When reconstructing these simulated measurements, we can compare the images with the underlying change in conductivity $\Delta\boldsymbol{\sigma}$ which serves as a well-defined reference to assess the reconstruction quality. This kind of evaluation would not be possible with real measurements, as there is no clearly defined ground truth available which could serve to assess the reconstruction quality. \Fref{fig:scenario_methods} illustrates the different steps of our procedure to quantify the quality of a certain reconstruction. In the following section, we will go through the different steps in more detail.\\
First, let us consider a pathological ventilation scenario, where only some regions of the lung lobes are ventilated while some others are not ventilated (\Fref{fig:scenario_methods}, Step 1). We define a reference state $\boldsymbol{\sigma}_\mathrm{ref}$ and an inspirated state $\boldsymbol{\sigma}_\mathrm{insp}$, where $\Delta\boldsymbol{\sigma} = \boldsymbol{\sigma}_\mathrm{insp} - \boldsymbol{\sigma}_\mathrm{ref}$ describes the conductivity change (\Fref{fig:scenario_methods}, Step 2). In our simulated ventilation scenario, we consider ventilated areas, that we also refer to as \textit{active},  with a conductivity change $\Delta\sigma_\mathrm{active}$ and, on the other hand, \textit{passive} areas without ventilation, i.e. without conductivity change. $\Delta\sigma_\mathrm{active}$ is defined as
\begin{align}
    \Delta\sigma_\mathrm{active} = \sigma_\mathrm{lung, aerated} - \sigma_\mathrm{lung, non-aerated}, 
\end{align}
where $\sigma_\mathrm{lung, areated}$ and $\sigma_\mathrm{lung, non-areated}$ are the conductivities of aerated and non-aerated lung tissue, respectively. The conductivity of the thorax $\sigma_{\mathrm{thorax}}$ remains constant. Note that in every sub-region of the geometry, the conductivity is uniform. Evaluating the forward model $F$ for the reference and inspirated state results in the simulated voltages $\brm{v}_\mathrm{ref}$ and $\brm{v}_\mathrm{insp}$, respectively (\Fref{fig:scenario_methods}, Step 3). After calculating the voltage difference $\Delta \brm{v}$ (\Fref{fig:scenario_methods}, Step 4), we add $n_{n}$ different noise samples $\brm{n}_i \in \mathbb{R}^{n_M}$ to $\Delta \brm{v}$ and reconstruct the corresponding images (\Fref{fig:scenario_methods}, Step 5). The purpose of distorting the voltage data with noise is to investigate the reconstruction quality not only in terms of accuracy but also in terms of robustness with respect to noise. To maintain the generalizability of our results, the noise samples are instances of an uncorrelated Gaussian noise with a certain amplitude. We choose it such that the voltage signal of the measurement channels reaches a signal-to-noise ratio (SNR), which is in the range of SNR in clinical EIT measurements. Using a reconstruction matrix $\brm{R}$, we then obtain a noisy image $\hat{\brm{x}}_i$ for every noise sample $i$ as well as the average image $\hat{\brm{x}}_m$ (\Fref{fig:scenario_methods}, Steps 5 and 6). 
\subsection{Measures of Reconstruction Quality}\label{sec:quality_measures}
To assess the reconstruction quality of different parametrizations of GREIT, we define measures for the accuracy and robustness of a certain reconstruction setup. For the definition of these quality measures, we first recall the reconstructed images $\hat{\brm{x}}_i$ as introduced in \Fref{fig:scenario_methods}, Step 5 and combine them in one image matrix $\hat{\brm{x}}$ as
\begin{align}
\hat{\brm{x}} =  [\hat{\brm{x}}_1, \hat{\brm{x}}_2, \ldots, \hat{\brm{x}}_{n_n} ]^\mathrm{T} = \begin{bmatrix} 
    \hat{x}_{11} & \dots  & \hat{x}_{1n_p} \\
    \vdots       & \ddots & \vdots        \\
    \hat{x}_{n_n1} & \dots  & \hat{x}_{n_n n_p} 
\end{bmatrix},
\end{align}
where the row represents the image index $i \in \{1,2,\ldots, n_n\}$ and the column the element (i.e. pixel) index $j \in \{1,2,\ldots, n_p\}$.
The average image of a series of noise samples $\hat{\brm{x}}_\mathrm{m}$ is then given by
\begin{align}
    \hat{\brm{x}}_\mathrm{m}=\frac{1}{n_n}\sum\limits_{i=1}^{n_n}\hat{\brm{x}}_i.
\end{align}
We employ two measures of reconstruction quality: The signal-to-noise ratio of the reconstructed images $\mathrm{SNR}_\mathrm{im}$ and the Pearson correlation coefficient $r_{\Delta\boldsymbol{\sigma}}$ with respect to the conductivity change $\Delta\boldsymbol{\sigma} = \boldsymbol{\sigma}_\mathrm{insp} - \boldsymbol{\sigma}_\mathrm{ref}$. Note that $\mathrm{SNR}_\mathrm{im}$ is the signal-to-noise ratio of the reconstructed images $\hat{\brm{x}}$ and is thus not equivalent to the $\mathrm{SNR}$ of the voltage data. For the definition of $\mathrm{SNR}_\mathrm{im}$ we use a simplified notation for the mean and standard deviation iterated over the row-index $i$ given by
\begin{align}
    &\mathrm{mean}_i (\hat{x}_{ij}) = \frac{1}{n_n}\sum\limits_{i=1}^{n_n}\hat{x}_{ij}, \\ 
    &\mathrm{std}_i (\hat{x}_{ij}) = \left[\frac{1}{n_n}\sum\limits_{i=1}^{n_n}(\hat{x}_{ij} -\mathrm{mean}_i (\hat{x}_{ij}))^2\right]^{0.5}.
\end{align}
Now $\mathrm{SNR}_\mathrm{im}$ can be written as the average of the pixel-wise $\mathrm{SNR}_j$ as 
\begin{align}
    \mathrm{SNR}_\mathrm{im} = \frac{1}{n_p} \sum\limits_{j=1}^{n_p} \mathrm{SNR}_{j} = \frac{1}{n_p} \sum\limits_{j=1}^{n_p} \frac{\mathrm{mean}_i(\hat{\mathrm{x}}_{ij})}{\mathrm{std}_i(\hat{\mathrm{x}}_{ij})}
\end{align}
Concluding this, we define our measure for noise on the image side $\mathrm{SNR}_\mathrm{im}$ differently compared to the commonly used noise figure $\mathrm{NF}$. In \citeasnoun{Adler2009}, $\mathrm{NF}$ is defined based on the noise amplification of the reconstruction of a single small non-conductive target in the center of the domain. In contrast to that, our definition of $\mathrm{SNR}_\mathrm{im}$ based on the simulated ventilation scenarios is closer to the real-world application in lung EIT and, therefore, more suitable for assessing the noise in a lung EIT image.\\
As mentioned above, the second measure of reconstruction quality is the correlation with respect to the conductivity change image $r_{\Delta\boldsymbol{\sigma}}$ which we define using the covariance $\mathrm{cov}(\cdot)$ as 
\begin{align}
    r_{\Delta\boldsymbol{\sigma}} = \frac{\mathrm{cov}(\hat{\brm{x}}_\mathrm{m},\,\Delta\boldsymbol{\sigma})}{\mathrm{std}(\hat{\brm{x}}_\mathrm{m})\,\mathrm{std}(\Delta\boldsymbol{\sigma})}.
\end{align}
It is employed as a measure of the accuracy of the reconstruction approach. Note that $r_{\Delta\boldsymbol{\sigma}} =1$ indicates a perfect reconstruction, which is not possible by definition. The diffusive and ill-posed character of the EIT problem prevents it. Nevertheless, the conductivity change correlation $r_{\Delta\boldsymbol{\sigma}}$ gives insight into how accurate the reconstruction is in terms of amplitude and shape errors.
\section{Results}
\subsection{Simulation Setup}\label{sec:scenario_setup_results}
\begin{figure}
    \centering
    \fontsize{10pt}{11pt}\selectfont
    \def\svgwidth{6in}
    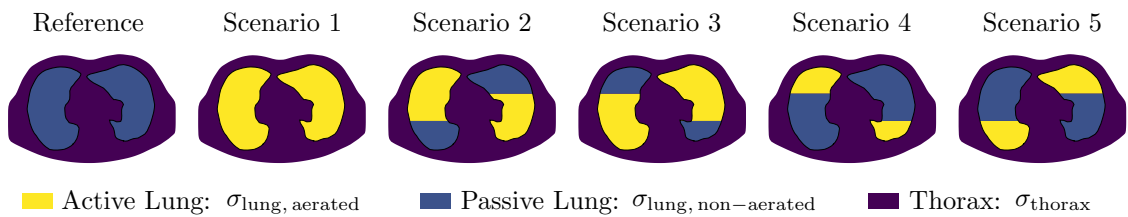
    \caption{Simulated ventilation scenarios with regionally homogeneous conductivities. Only the active regions of the lungs are subject to a conductivity change.}
    \label{fig:sim_scenarios}
\end{figure}

As already introduced in Sections \ref{sec:scenario_setup_method} and \ref{sec:quality_measures}, we perform the following investigations of the reconstruction parameter space based on artificial ventilation scenarios as this has the advantage of a well-defined ground truth. See \Fref{fig:scenario_methods} for a summary of the general procedure. All necessary parameters to generate our simulated noisy electrode voltages can be found in Table \ref{tab:params_sim_setup}. \\
We consider five different ventilation scenarios, as illustrated in \Fref{fig:sim_scenarios}. In each scenario, two non-contiguous regions are subject to a conductivity change of $\Delta \sigma_\mathrm{active}=\sigma_\mathrm{lung,aerated} - \sigma_\mathrm{lung,non-aerated}$. In the passive regions of the domain, the conductivity remains constant. The scenarios are a simplified representation of a highly heterogeneous lung where only parts of the lung lobes are ventilated. Only in scenario 1 the entire lung masks are active and thus subject to a conductivity change. To avoid misinterpretation of the reconstructions, it is essential that our reconstruction setup reproduces both active regions with the same amplitude.\\
Motivated by the SenTec SensorBelt (SenTec, Landquart, Switzerland)\cite{Sophocleous2018}, we set up a thorax model that uses 32 rectangular electrodes that are paired with a \textit{skip4} pattern for stimulation and measurement. At the stimulating electrodes, a current of $I_\mathrm{stim} = 5\,\mathrm{mA}$ is applied. The model geometry is created by extruding a transverse slice of the CT image of one exemplary ARDS patient. Thus, the model does not reflect any changes in the patient's longitudinal axis. The geometrical model, including dimensions and mesh-related parameters, is shown in the Supplement S.1. As introduced in Section \ref{sec:scenario_setup_method}, we add $n_n=20$ uncorrelated Gaussian noise samples $\brm{n}_i$ to the measurement channels such that we achieve a desired average $\mathrm{SNR}$. Motivated by clinical measurements, we set this value to $\mathrm{SNR} = 7.2\, \mathrm{dB}$.\\
Table \ref{tab:overview_GREIT_params} lists all relevant parameters for the GREIT reconstructions shown in the Sections \ref{sec:param_study_unif} and \ref{sec:param_study_cust}.
\begin{table}[b]
    \centering
    \begin{tabular}{|p{3cm} | p{1.5cm} || p{3cm} | p{1.5cm}|}
        \hline
        $\sigma_\mathrm{lung, aerated}$     & $0.06\,\frac{1}{\Omega \mathrm{m}}$  & $\mathrm{SNR}$ & $7.2\,\mathrm{dB}$ \\
        $\sigma_\mathrm{lung,non-aerated}$     & $0.12\,\frac{1}{\Omega \mathrm{m}}$ & $I_{\mathrm{stim}}$ & $5\,\mathrm{mA}$ \\
        $\sigma_\mathrm{thorax}$     & $0.48\,\frac{1}{\Omega \mathrm{m}}$ & $n_\mathrm{elec}$ & $32$ \\
        $n_n$ & 20 & & \\
        \hline
    \end{tabular}
    \caption{Parameters to simulate noisy electrode voltages. The values for the tissue conductivities are taken from \protect\cite{Adler1996a}.}
    \label{tab:params_sim_setup}
\end{table}%
\subsection{Influence of Background Conductivity on Noise Performance} \label{sec:param_study_unif}
\begin{figure}
    \centering
    \fontsize{10pt}{11pt}\selectfont
    \def\svgwidth{6in}
    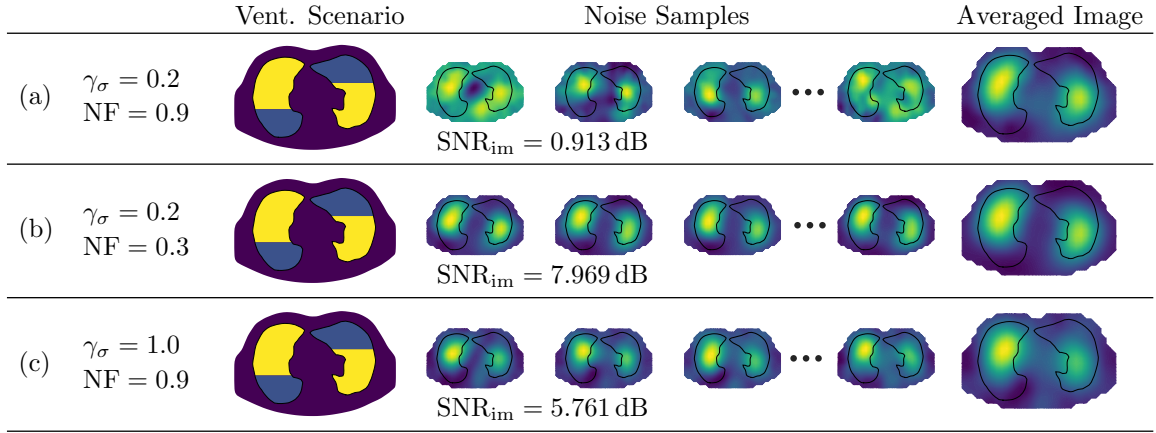
    \caption{Three parameterization samples of a reconstruction setup with uniform target distribution. The figure shows exemplarily the results of scenario 2. $\gamma_\sigma$ and $\mathrm{NF}$ represent the background conductivity ratio as introduced in Equation \eqref{eq:background_conductivity_ratio} and the noise figure, respectively.}
    \label{fig:noise_samples_unif}
\end{figure}

\begin{figure}
    \centering
    \fontsize{10pt}{11pt}\selectfont
    \def\svgwidth{6in}
    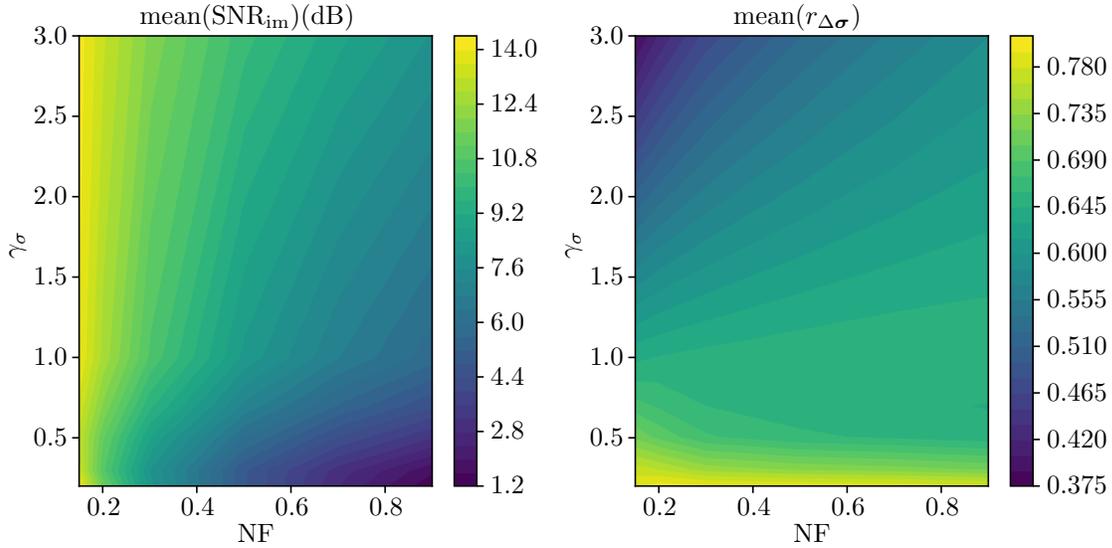
    \caption{Evaluation of the parameter space for a uniform target distribution in terms of $\mathrm{SNR_{im}}$ and the correlation coefficient with respect to the conductivity change image $r_{\Delta\boldsymbol{\sigma}}$. Both quality measures are averaged over the five simulation scenarios.}
    \label{fig:unif_param_space}
\end{figure}

First, we show the effect of the distribution of the background conductivity by varying the ratio $\gamma_\sigma = {\sigma_{\mathrm{ref,\,lung}}}/{\sigma_{\mathrm{ref,\,thorax}}}$ as introduced in Section \ref{sec:background_conductivity}. The distribution of training targets is kept uniform, and the $\mathrm{NF}$ is varied within the range of 0.15 to 0.9. See all reconstruction parameters in Table \ref{tab:overview_GREIT_params}.\\
\begin{table}[]
    \centering
    \begin{tabular}{l l l}
    \hline
         Parameter   & Uniform target distribution & Customized target distribution \\ 
         & Section \ref{sec:param_study_unif} & Section \ref{sec:param_study_cust}\\
        \hline
        $N_t$ & 1000   & 1000   \\
        $s_{targ}$ & 0.02   & 0.02   \\
        $R_w$ & 0.2 & $\{0.1 , 0.13, 0.17, 0.2\}$  \\
        $\mathrm{NF}$ & $\{0.15, 0.2, 0.3, 0.5, 0.7, 0.9\}$ & 0.3 \\
        $\gamma_\sigma$ & $\{0.1, 0.2, 0,3, 0.5, 0.7, 1.0, 3.0\}$ & 0.2 \\
        Targ. Distr. & uniform, GREIT \#3 & regionally uniform \\
        $\gamma_\rho$ & - & $\{1.0, 1.5, 2.0, 2.5, 3.0, 3.5\}$ \\
        $s_{bound} $ & - & $2\,\mathrm{mm}$ \\
        \hline
    \end{tabular}
    \caption{GREIT parameter setting for the uniform and customized target distributions in Sections \ref{sec:param_study_unif} and \ref{sec:param_study_cust}.}
    \label{tab:overview_GREIT_params}
\end{table}%
\Fref{fig:noise_samples_unif} presents three exemplary parameterizations (a) - (c) for the ventilation Scenario 2. For each parameterization, we show four exemplary reconstructions of the noise instances as well as the averaged image $ \hat{\brm{x}}_\mathrm{m}$ over all 20 noise samples. As shown in  \Fref{fig:noise_samples_unif}  (a) and (b), the noise level of the reconstructed image series strongly depends on the chosen value for $\mathrm{NF}$. So far, this is obvious, as $\mathrm{NF}$ is the main tool to adapt noise performance and thus the regularization level in GREIT. Another observation, however, is not as intuitive. Leaving $\mathrm{NF}$ constant and manipulating the conductivity ratio $\gamma_\sigma$ also significantly affects the noise performance of GREIT, as seen in \Fref{fig:noise_samples_unif} (c). Higher $\gamma_\sigma$ reduce the influence of noise, leading to increased $\mathrm{SNR}_{\mathrm{im}}$. However, when comparing the ground truth ventilation scenario to the averaged reconstruction in \Fref{fig:noise_samples_unif} (c), it becomes apparent that both lung regions do not show the same intensity in the EIT image, although this would be expected due to the same amplitude of the conductivity change in both lung regions.\\
To investigate the influence of $\gamma_\sigma$ and $\mathrm{NF}$ for a broader range of parameter combinations, \Fref{fig:unif_param_space} shows the noise performance by means of the $\mathrm{SNR}_{\mathrm{im}}$ as well as the correlation with respect to the conductivity change $r_{\Delta\boldsymbol{\sigma}}$ averaged over all ventilation scenarios. The observation that the noise performance of GREIT improves with decreasing $\mathrm{NF}$ and increasing background conductivity ratio is confirmed by \Fref{fig:unif_param_space}. Especially in the region $\gamma_\sigma<1.0$, the noise performance shows the highest gradient in the direction of $\gamma_\sigma$, meaning that a more physiological choice of the background conductivity comes with the price of a poor noise performance. This effect can only partially be compensated by the noise figure.  Additionally, it becomes apparent that the correlation with ground truth is not severely influenced by $\mathrm{NF}$ but strongly deteriorates when increasing $\gamma_\sigma$.
\subsection{Noise Performance and Accuracy of Customized Target Distributions}\label{sec:param_study_cust}
\begin{figure}
    \centering
    \fontsize{10pt}{11pt}\selectfont
    \def\svgwidth{6in}
    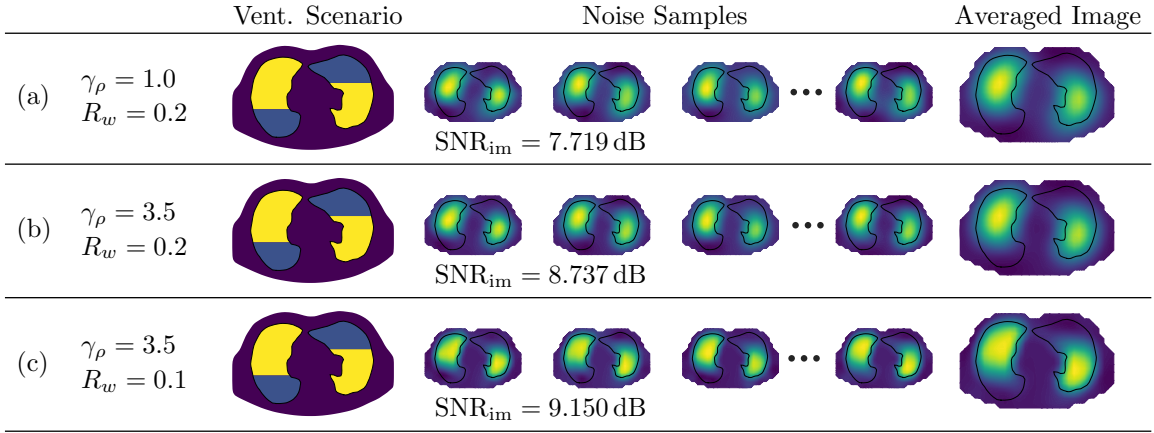
    \caption{Three parameterization samples of a reconstruction setup with custom target distribution. The figure shows exemplarily the results of scenario 2. $\gamma_\rho$ and $R_w$ represent the training target density ratio as defined in Equation \eqref{eq:target_density_ratio} and the weighting radius as introduced in Section \ref{sec:GREIT_fundamental}, respectively.}
    \label{fig:noise_samples_cust}
\end{figure}

\begin{figure}
    \centering
    \fontsize{10pt}{11pt}\selectfont
    \def\svgwidth{6in}
    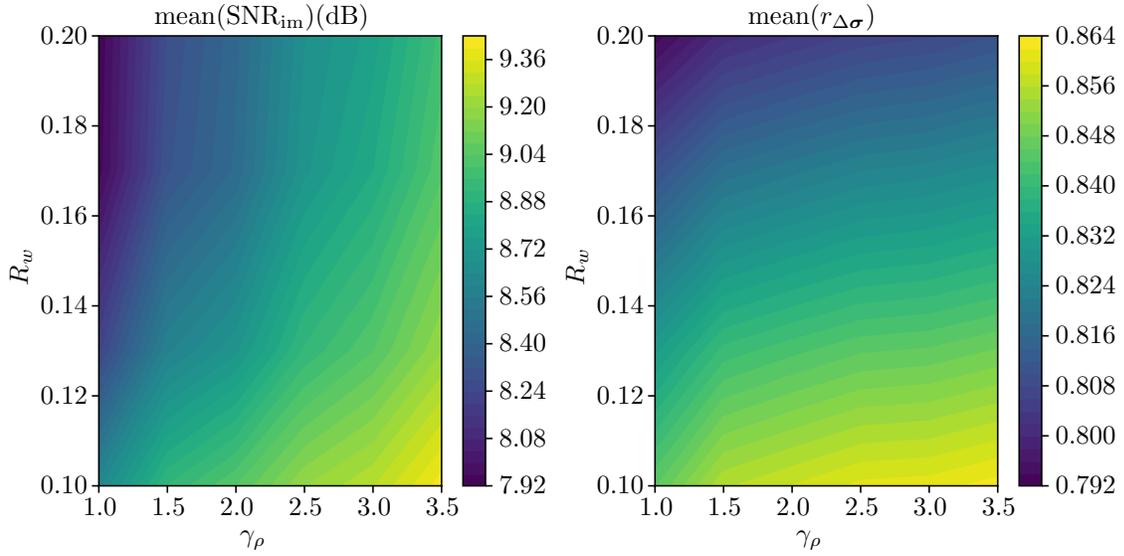
    \caption{Evaluation of the parameter space for a customized target distribution in terms of $\mathrm{SNR_{im}}$ and the correlation coefficient with respect to the conductivity change image $r_{\Delta\boldsymbol{\sigma}}$. Both quality measures are averaged over the five simulation scenarios.}
    \label{fig:cust_param_space}
\end{figure}

As introduced in Section \ref{sec:custom_target_distr}, manipulating the distribution of non-conductive training targets allows the incorporation of anatomical information of the thorax and lung geometry into the reconstruction. For the following investigations on the distribution of training targets, we set the noise figure and background conductivity ratio constant to $\mathrm{NF} = 0.3$ and $\gamma_\sigma=0.2$. Also, the minimal distance to the thorax boundary $s_{bound}$ is kept fixed as no significant influence on the evaluated reconstruction quality measures was observed. All other parameters are listed in Table \ref{tab:overview_GREIT_params}. 
\Fref{fig:noise_samples_cust} shows the reconstructed images of the simulation scenario 2 for three exemplary parameter combinations (a) - (c) of the target density ratio $\gamma_\rho$ and the weighting radius $R_w$. Comparing parameterizations (a) - (c) shows that modest improvements in terms of noise performance can be reached by increasing $\gamma_\rho$ and reducing $R_w$. This is also emphasized by the plot of $\mathrm{SNR_{im}}$ over the entire parameter space and all simulation scenarios in \Fref{fig:cust_param_space}.  Furthermore, the adapted target distribution slightly reduces the amount of signal outside the lungs. Particularly, for the average image in \Fref{fig:noise_samples_cust} (c), the signal is almost entirely confined within the lung mask. This increases the correlation with the conductivity change image $r_{\Delta\boldsymbol{\sigma}}$ as shown in \Fref{fig:cust_param_space}. \Fref{fig:noise_samples_cust} (c) also indicates that reducing the weighting radius to $R_w=0.1$ decreases the asymmetry between the left and right maximum compared to (a) and (b).\\
\\
\subsection{Application of the Adapted Parametrization on Clinical Data from an ARDS Patient}\label{sec:showcase_e9}
To highlight how the clinical validity of EIT images is influenced by the reconstruction itself and how it can be improved by adjusting the parameters and target distribution, we compare the EIT reconstructions of three different parameterizations applied to clinical data from an exemplary ARDS patient. The clinical data were collected at the Department of Anesthesiology and Intensive Care Medicine at the University Medical Center Schleswig-Holstein in Kiel (Germany) as part of the SMART study (DRKS-ID: DRKS00017151) on the recruitment/de-recruitment behavior of lung tissue in ARDS patients. We performed the following investigations based on the geometry of one of these patients shown in the Supplement S.2. \\
\Fref{fig:showcase_e9} shows the CT image and the EIT reconstructions for the considered patient. We compare the EIT results of three different reconstruction setups. $\brm{R}_\mathrm{A}$ is equivalent to the parameter choice as it is made in the tutorials coming with EIDORS\footnote[7]{\url{https://eidors3d.sourceforge.net/tutorial/GREIT/adult_ex.shtml}}, $\brm{R}_\mathrm{B}$ to the optimal parameter set as proposed by \citeasnoun{Thurk2017} and $\brm{R}_\mathrm{C}$ is the reconstruction setup with customized parametrization as proposed in our work. The parameter settings are shown in Table \ref{tab:showcase_e9_params}. \\
When looking at the different parametrizations of GREIT in \Fref{fig:showcase_e9}, we see that all three setups predict quite well the location of the consolidations in the dorsal regions of the patient's lungs. Comparing $\brm{R}_\mathrm{A}$ and $\brm{R}_\mathrm{B}$, we already observe an increase in anatomical accuracy associated with a decrease in the weighting radius. It also becomes apparent that informing the reconstruction with geometrical features (i.e., by including non-homogeneous conductivity and/or target distribution) increases the anatomical accuracy further. For $\brm{R}_\mathrm{C}$, most of the EIT signal is concentrated within the lung masks, which makes the images easier to interpret for clinical use. 

\begin{figure}
    \centering
    \fontsize{10pt}{11pt}\selectfont
    \def\svgwidth{5in}
    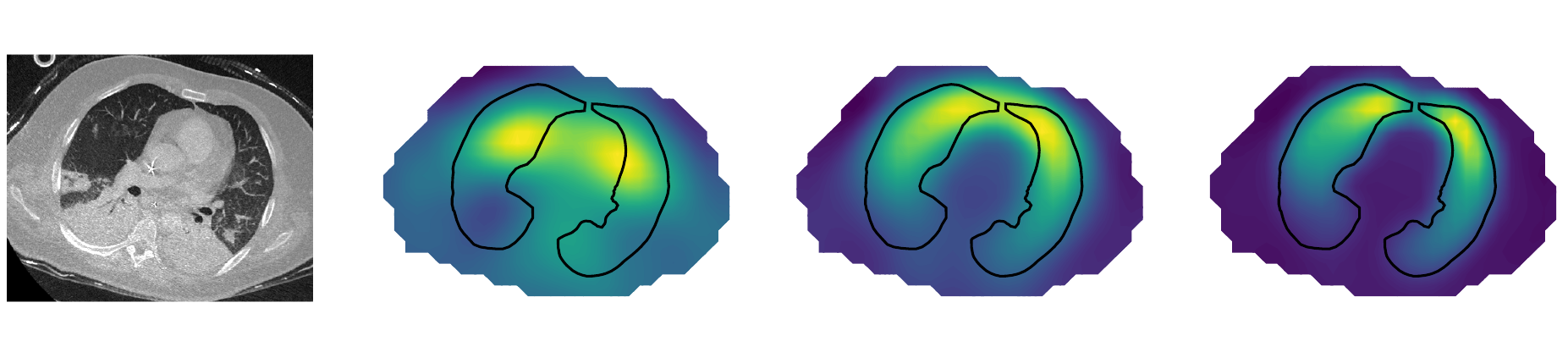
    \caption{Influence of the GREIT parameterization on the reconstructed images during mechanical ventilation of an exemplary ARDS patient. The CT image of an exemplary ARDS patient can be compared with three different parametrizations of GREIT. $\brm{R}_{\mathrm{A}}$ and $\brm{R}_{\mathrm{B}}$ represent reconstruction matrices as they have been used in literature, namely the EIDORS tutorials\protect\footnote[7]{} and \protect\citeasnoun{Thurk2017}, respectively. $\brm{R}_{\mathrm{C}}$ corresponds to a customized target distribution as it is proposed in the previous sections. }
    \label{fig:showcase_e9}
\end{figure} 
\begin{table}[]
    \centering
    \begin{tabular}{l l l l}
    \hline
          & $\brm{R}_\mathrm{A}$ & $\brm{R}_\mathrm{B}$ & $\brm{R}_\mathrm{C}$ \\ 
        \hline
        $N_t$   & 500 & 1000   & 1000   \\
        $s_{targ}$  & 0.03 & 0.06   & 0.02   \\
        $R_w$ & 0.25 & 0.15 & 0.1 \\
        $\mathrm{NF}$ & 0.5 &  0.15 & 0.15 \\
        $\gamma_{\sigma,\mathrm{lung}}$ & 1.0 & 0.2  & 0.2 \\
        $\gamma_{\sigma,\mathrm{heart}}$ & 1.0 & 1.5  & 1.0 \\
        Targ. Distr. & uniform, GREIT \#3 & uniform, GREIT \#3 &  regionally uniform \\
        $\gamma_\rho$  & - & - & 3.5 \\
        $s_{bound} $ & - & - & $2\,\mathrm{mm}$ \\
        \hline
    \end{tabular}
    \caption{GREIT reconstruction parameters of three different parameterizations $\brm{R}_\mathrm{A}$, $\brm{R}_\mathrm{B}$ and $\brm{R}_\mathrm{C}$ that are applied on the clinical showcase scenario in Section \ref{sec:showcase_e9}.}
    \label{tab:showcase_e9_params}
\end{table}%
\section{Discussion}
This study quantitatively investigates how the quality of EIT images reconstructed by GREIT is influenced by some of the parameters of the reconstruction algorithm, namely the background conductivity, the weighting radius, and the distribution of training targets. Image quality is assessed using the signal-to-noise ratio of the images, $\mathrm{SNR}_\mathrm{im}$, and the correlation coefficient with respect to the conductivity change, $r_{\Delta \sigma}$. We observed that the different parameters significantly influence the reconstructed images and that their interpretability can be improved by a sophisticated choice of reconstruction parameters. \\
\textit{Limitations:} The parameter studies presented here were conducted using simulated voltage data, which offers the advantage of having a known ground truth for evaluating reconstruction accuracy. However, this approach has inherent limitations, as simulated voltages differ from actual clinical measurements. One source of discrepancy is the inaccuracy of the forward model, which includes simplifications such as regionally homogeneous conductivities, a 2.5D thorax representation, and the omission of thoracic motion. Additionally, we assumed uncorrelated noise in the data, which is likely not reflective of real-world conditions.\\
Despite these limitations, we observed consistent results across five different ventilation scenarios and various noise levels, suggesting that our findings are transferrable to clinical data. Thus far, we have applied the optimized reconstruction setup to measurements from a single patient with heterogeneous lung pathology. This showcase scenario serves as a demonstration of feasibility rather than evidence of generalizability. Future work should aim to validate the influence of reconstruction parameters using e.g. 4D-CT data, as it was demonstrated by \citeasnoun{Katayama2024}. So far, we only validate the reconstruction approach using the established validation framework of \citeasnoun{Grychtol2014}, that is based on porcine models, see Appendix. \\
\textit{Background conductivity.} One parameter strongly influencing the reconstructed images is the background conductivity. For a broad range of $\mathrm{NF}$, the noise level on the image side is severely reduced with increasing ratio of the background conductivities $\gamma_\sigma$, as shown in \Fref{fig:unif_param_space}. Hence, in the case of high noise in clinical voltage data, increasing the ratio $\gamma_\sigma$ sounds appealing to reduce the artifacts of noise in EIT images. However, this means linearizing around a state far from the physiological reference, which comes at the price of significantly deteriorated accuracy. Consequently, our findings corroborate the observation of \citeasnoun{Grychtol2013}, who state that a uniform assumption for the background conductivity distribution produces misleading EIT images. The misleading character of EIT images based on unphysiological assumptions for the background conductivity also becomes apparent when comparing the amplitude of the two maxima that represent the lungs in our simulation cases, see \Fref{fig:noise_samples_unif}. Although both lung regions are subject to the same conductivity change in the forward model, the resulting reconstructed images imply a severely asymmetrical ventilation of the lungs. In clinical practice, such misinterpreted evidence might lead to potentially harmful therapeutic decisions of mechanically ventilated patients. To further reduce the risk of producing misleading EIT images, one could also consider incorporating the heterogeneous nature of ARDS lungs by locally adapting the background conductivity to the Hounsfield units. \\
\textit{Weighting radius.}
In the second step, we varied the weighting radius $R_w$ of the non-conductive training targets used in GREIT. $R_w$ can also be interpreted as the desired radius that the image of a spherical training target should have. Therefore, a reduction of $R_w$ increases the resolution of the reconstructed images. In the simulation scenarios considered to assess the reconstruction quality, the increased resolution with decreasing $R_w$ is reflected in an increasing correlation with respect to the conductivity change $r_{\Delta\sigma}$, see \Fref{fig:cust_param_space}. Additionally, the asymmetry of the two lung maxima, which is described above for unphysiological assumptions of the background conductivity, is reduced by a smaller weighting radius (\Fref{fig:noise_samples_cust}). Concluding this, we propose the choice of \textit{small} values for the weighting radius $R_w$. The definition of \textit{small} depends on the choice of the initial target size $s_{targ}$. In the present work, we only varied $R_w$, i.e., the size of the desired images of the training targets. The size of the targets themselves was not changed. Presumably, it is the interplay between both, the weighting radius $R_w$ and the target size $s_{targ}$, that influences the image resolution and accuracy. Additionally, one could argue that the higher the number of electrodes, the smaller the desired target radius should be, as the electrode distance interacts with the image resolution. For a 16-electrode setup, \citeasnoun{Adler2009} recommended a $R_w$ of $0.2$ times the medium diameter. For a setup with 32 electrodes, the electrode distance halves. Hence, we propose $R_w = 0.1$.\\
\textit{Customized target distribution.} This work also presents reconstruction results based on an adapted spatial distribution of the non-conductive training targets. By concentrating more targets within the lung mask, we were able to reduce the image signal outside the lungs and mitigate the effects of measurement noise, leading to images that better correspond to the lung anatomy and thereby improve the interpretability of EIT images. However, this approach needs to be applied with care, as the inhomogeneous distribution of targets clearly introduces some bias. The effects of an impedance change in regions of low target density (i.e. the thorax) are more likely to be interpreted as an impedance change inside regions of high target density (i.e., the lungs). This carries the inherent risk of misinterpretation. Hence, we suggest applying a customized target distribution in parallel to a regular reconstruction that is based on a uniform target density. Thereby, one can assess if an impedance change outside the lungs (e.g. induced by the heart beat) might be moved into the lung mask by the high target density and hence be misinterpreted as a ventilation-related impedance change. 

\section{Conclusion}
In this study, we investigated the impact of different parameters on the quality of reconstructed EIT images using the well-established GREIT algorithm. To quantitatively assess the reconstruction quality, we present a methodology that define quality measures based on simulated and physiologically motivated scenarios. Our findings highlight that the choice of reconstruction parameters significantly influences the EIT images. Specifically, we found that an unphysiological assumption for the GREIT background conductivity produces significantly misleading EIT images but simultaneously reduces the influence of measurement noise. Additionally, we applied an adapted training target distribution where the spatial density of the targets is higher within the lung regions. By increasing the target density inside the lungs and decreasing the desired size of the target images (i.e., the weighting radius), we were able to increase the resolution and anatomical accuracy. We also demonstrated for an exemplary ARDS patient how the new parameterization of GREIT would affect clinical reconstructions and how EIT images would benefit in terms of anatomical interpretability.
\section*{Acknowledgments}
The authors gratefully acknowledge financial support by BREATHE, a Horizon 2020—ERC–2020–ADG project (grant agreement No. 101021526-BREATHE), and by the Deutsche Forschungsgemeinschaft (DFG, German Research Foundation) in the projects WA1521/26-1 and BE6526/1-1.
\section*{Ethical statement}
In this study we analyzed data collected in the SMART-Study (DRKS-ID: DRKS00017151). This study was approved by the Ethics Committee of the Medical Faculty of the Christian-Albrechts-Universität Kiel (Ethics Committee Number D 485/17) and carried out in accordance with the Helsinki Declaration. Written informed consent was obtained from all patients or their legal representatives.
\section*{Data availability}
The data that support the findings of this study are available on request from the corresponding author. The data are not publicly available due to privacy or ethical restrictions.
\clearpage
\newpage

\section*{Appendix A}\label{sec:appendix}
\subsection*{Validating the Ability to Reproduce Physiological EIT Measures}\label{sec:physiological_validation}
\begin{table}[b]
    \centering
    \begin{tabular}{l l l}
    \hline
         Parameter  & $\brm{R}_{\mathrm{unif}}$ & $\brm{R}_{\mathrm{cust}}$ \\ 
        \hline
        $N_t$   & 1000   & 1000   \\
        $s_{targ}$  & 0.02   & 0.02   \\
        $R_w$ & 0.2 & 0.1 \\
        $\mathrm{NF}$ & 0.3 & 0.3 \\
        $\gamma_\sigma$ & 0.2 & 0.2 \\
        Targ. Distr. & uniform, GREIT \#3 & regionally uniform \\
        $\gamma_\rho$  & - & 3.5 \\
        $s_{bound} $ & - & $0\,\mathrm{mm}$ \\
        \hline
    \end{tabular}
    \caption{GREIT parameter setting for the uniform and customized reconstruction matrices $\mathrm{\mathbf{R}_{\mathrm{unif}}}$ and $\mathrm{\mathbf{R}_{\mathrm{cust}}}$, respectively.}
    \label{tab:pig_valid_params}
\end{table}%
As already introduced in the main part of this study, we also validated the reconstruction matrices obtained from our adapted parameter set by applying the validation workflow of \citeasnoun{Grychtol2014}. Based on data obtained from eight mechanically ventilated pigs, they propose a methodology to assess the eligibility of a reconstruction method to reproduce functional EIT measures correctly. We apply this approach to the reconstruction setup that is best suitable for the ventilation scenarios described in the main part of this work.  Therefore, we make use of the sample EIDORS pig thorax geometry\footnote[8]{\url{https://eidors3d.sourceforge.net/tutorial/netgen/extrusion/pig_body.shtml}}. The validation procedure is briefly described below. For further details, the reader is referred to the original publication \cite{Grychtol2014}. \\
\begin{table}[]
\centering
\begin{tblr}{
  colspec = {|l|Q[h,c]Q[h,c]||Q[h,c]Q[h,c,wd=2cm]Q[h,c,wd=2cm]|},
  row{4}={bg=table-gray},
  row{7}={bg=table-gray},
  row{10}={bg=table-gray},
  row{13}={bg=table-gray},
  row{16}={bg=table-gray},
  row{19}={bg=table-gray}
  }
\hline
 & $\mathrm{\mathbf{R}_{\mathrm{unif}}}$ & $\mathrm{\mathbf{R}_{\mathrm{cust}}}$ & $\mathrm{\mathbf{R}_{\mathrm{SBP}}}$ & $\mathrm{\mathbf{R}_{\text{GR,Grychtol}}}$ & $\mathrm{\mathbf{R}_{\text{GR,Th\"urk}}}$ \\
\hline
$\brm{PEEP}= \brm{0\,cm\,H_2O}$& \includegraphics[height=1cm]{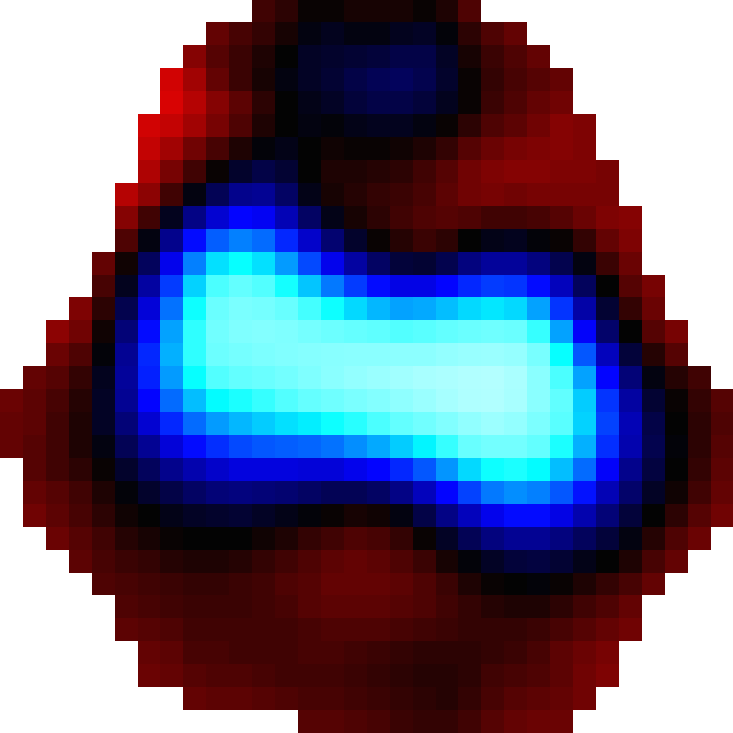} & \includegraphics[height=1cm]{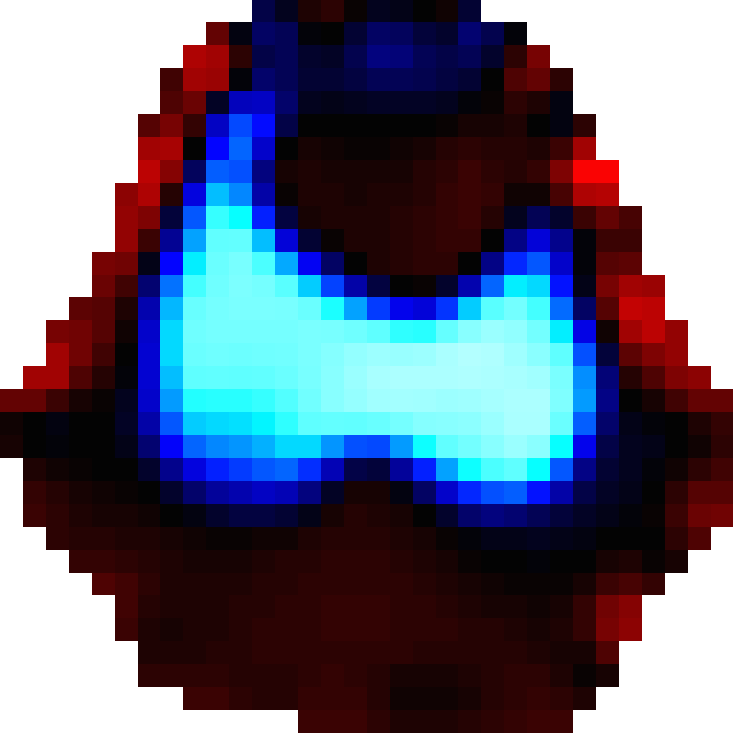} & \includegraphics[height=1cm]{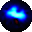} & \includegraphics[height=1cm]{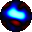} & \includegraphics[height=1cm]{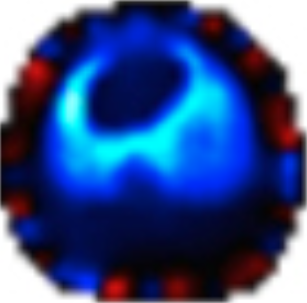}\\
$\brm{PEEP}= \brm{5\,cm\,H_2O}$ & \includegraphics[height=1cm]{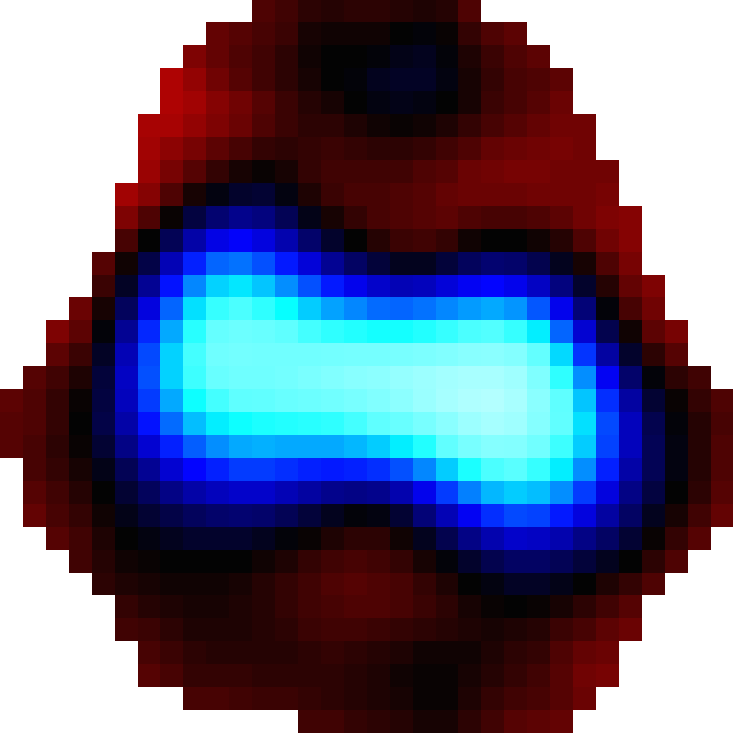} & \includegraphics[height=1cm]{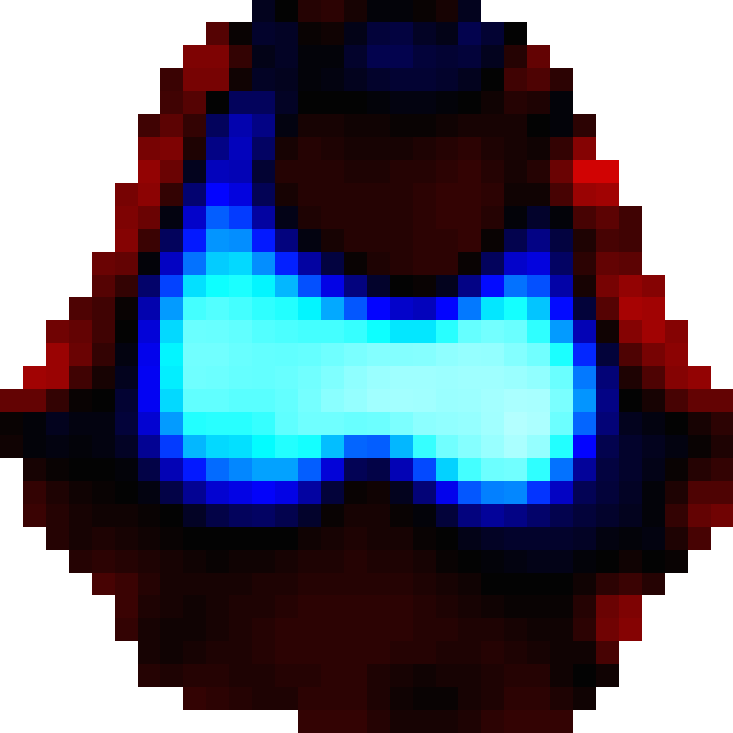} & \includegraphics[height=1cm]{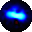}& \includegraphics[height=1cm]{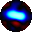} & \includegraphics[height=1cm]{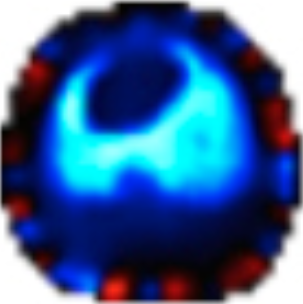}\\
\hline
TV is independent of PEEP &  &  &  &  & \\
\quad$\mathrm{TV_{E,Z,21}} = \mathrm{TV_{E,P,21}}$      & $0.000^*$ & $0.000^*$ & $0.044^*$ & $0.018^*$ & $0.019^*$ \\
\quad$\mathrm{TV_{E,Z,100}} = \mathrm{TV_{E,P,100}}$    & $0.001^*$ & $0.002^*$ & $0.013^*$ & $0.010^*$ & $0.007^*$ \\
TV is independent of $\mathrm{FiO_2}$ &  &  &  &  & \\
\quad$\mathrm{TV_{E,Z,21}} = \mathrm{TV_{E,Z,100}}$     & $0.054\;\,$   & $0.021^*$ & $0.004^*$ & $0.010^*$ & $0.009^*$ \\
\quad$\mathrm{TV_{E,P,21}} = \mathrm{TV_{E,P,100}}$     & $0.000^*$ & $0.000^*$ & $0.001^*$ & $0.002^*$ & $0.001^*$ \\
TV is reproducible &  &  &  &  & \\
\quad$\mathrm{TV_{E,Z1}} = \mathrm{TV_{E,Z2}}$          & $0.000^*$ & $0.000^*$ & $0.001^*$ & $0.001^*$ & $0.001^*$ \\
\quad$\mathrm{TV_{E,P1}} = \mathrm{TV_{E,P2}}$          & $0.000^*$ & $0.000^*$ & $0.000^*$ & $0.001^*$ & $0.000^*$ \\
CoV is PEEP dependent &  &  &  &  &  \\ 
\quad$\mathrm{CoV_{E,Z,21}} > \mathrm{CoV_{E,P,21}}$    & $0.023^*$ & $0.029^*$ & $0.064\;\,$   & $0.043^*$ & $0.036^*$ \\
\quad$\mathrm{CoV_{E,Z,100}} > \mathrm{CoV_{E,P,100}}$  & $0.013^*$ & $0.012^*$ & $0.023^*$ & $0.019^*$ & $0.013^*$ \\
CoV is $\mathrm{FiO_2}$ dependent &  &  &  &  &  \\ 
\quad$\mathrm{CoV_{E,Z,21}} < \mathrm{CoV_{E,Z,100}}$   & $0.013^*$ & $0.006^*$ & $0.010^*$ & $0.011^*$ & $0.007^*$ \\
\quad$\mathrm{CoV_{E,P,21}} < \mathrm{CoV_{E,P,100}}$   & $0.657\;\,$   & $0.608\;\,$   & $0.683\;\,$   & $0.604\;\,$   & $0.622\;\,$   \\
CoV is reproducible &  &  &  &  &  \\
\quad$\mathrm{CoV_{E,Z1}} = \mathrm{CoV_{E,Z2}}$        & $0.016^*$ & $0.005^*$ & $0.007^*$ & $0.032^*$ & $0.011^*$ \\
\quad$\mathrm{CoV_{E,P1}} = \mathrm{CoV_{E,P2}}$        & $0.003^*$ & $0.001^*$ & $0.002^*$ & $0.017^*$ & $0.001^*$ \\
\hline
\end{tblr}
\caption{Functional analysis of the reconstruction setup with respect to expected physiological behavior, as first proposed by \protect\citeasnoun{Grychtol2014}.  We compare the results of our reconstruction setups $\brm{R}_{\mathrm{unif}}$ and $\brm{R}_\mathrm{cust}$ with those from the original publication - $\brm{R}_\mathrm{SBP}$ and $\brm{R}_\mathrm{GR,Grychtol}$ for Sheffield backprojection and Grychtol's GREIT parameterization - and the optimal GREIT parameterization of \protect\citeasnoun{Thurk2019} $\brm{R}_\text{GR,Th\"urk}$. The EIT images in the first two rows correspond to $\text{PEEP}$ and $\text{ZEEP}$ of one exemplary animal (pig \#5). For each hypothesis, the $p$-values of the Student’s $t$-test are reported. For each test, $N=8$ (the number of animals). $\mathrm{E}$, measures by EIT; $\mathrm{TV}$, tidal volume; $\mathrm{PEEP}\,(\mathrm{P})$, positive end-expiratory pressure; $\mathrm{ZEEP}\,(\mathrm{Z})$, zero end-expiratory pressure; $\mathrm{FiO_2}$, fraction of $\mathrm{O}_2$ in inspired gas; 21, $\mathrm{FiO_2}$ equal to 21\%; 100, $\mathrm{FiO_2}$ equal to 100\%; indices 1 and 2 identify the first and the second measurements at identical $\mathrm{ZEEP}$ or $\mathrm{PEEP}$ levels; $\mathrm{CoV}$, center of ventilation; $^*$, hypothesis confirmed at $p\leq 0.05$.}
\label{tab:physiological_validation}
\end{table}%
During the mechanical ventilation of the animal, \citeasnoun{Grychtol2014} controlled the level of positive end-expiratory pressure (PEEP), the tidal volume, and the $\mathrm{FiO_2}$ level and recorded EIT measurements with a 16-electrode belt. The reconstruction methods were then assessed based on whether they were suitable to confirm the physiological hypotheses summed up in Table \ref{tab:physiological_validation}. There, we report the corresponding $p$-values of the Student's $t$-test for the parameterization of a uniform and a customized target distribution, $\brm{R}_{\mathrm{unif}}$ and $\brm{R}_{\mathrm{cust}}$, respectively. With $\brm{R}_{\mathrm{unif}}$ we refer to the reconstructions in Section \ref{sec:param_study_unif}, where we employ a uniform target distribution with a physiologically motivated background conductivity. $\brm{R}_{\mathrm{cust}}$ corresponds to the findings in Section \ref{sec:param_study_cust}, where we additionally concentrate the training targets in the regions of particular interest, which are, in our case, the lungs. The respective parameters of $\brm{R}_{\mathrm{unif}}$ and $\brm{R}_{\mathrm{cust}}$ can be looked up in Table \ref{tab:pig_valid_params}. For reference, Table \ref{tab:physiological_validation} also contains the $p$-values of the Sheffield backprojection, the GREIT parameterization of \citeasnoun{Grychtol2014} and the optimal GREIT parameterization of \citeasnoun{Thurk2019}, $\brm{R}_{\mathrm{SBP}}$, $\brm{R}_{\mathrm{GR,Grychtol}}$ and $\brm{R}_\text{GR,Th\"urk}$, respectively. The results for these reference reconstructions are taken from the original publications\cite{Grychtol2014,Thurk2019}. Hypotheses with $p\leq 0.05$ are considered confirmed. \\
$\brm{R}_{\mathrm{unif}}$ and $\brm{R}_{\mathrm{cust}}$ can both confirm most of the hypothesis. The only hypothesis that is missed by $\brm{R}_{\mathrm{unif}}$ but confirmed by all others is the independence of the tidal volume $\mathrm{TV}$ on $\mathrm{FiO_2}$. Equivalent to the already established reconstruction setups,  $\brm{R}_{\mathrm{unif}}$ and $\brm{R}_{\mathrm{cust}}$ were also not capable of reproducing the dependence of the center of ventilation $\mathrm{CoV}$ on the $\mathrm{FiO_2}$ at $\mathrm{PEEP}$ level. This can be attributed to the absorption atelectasis - the mechanism why $\mathrm{CoV}$ changes with $\mathrm{FiO_2}$ -  that is reduced at higher PEEP levels \cite{Magnusson2003}. All other physiological hypotheses can be reproduced with our reconstruction matrices $\brm{R}_{\mathrm{unif}}$ and $\brm{R}_{\mathrm{cust}}$.

\newpage
\section*{References}
\bibliographystyle{dcu} 
\bibliography{literature}

\end{document}


\renewcommand{\thesection}{S.\arabic{section}}
\renewcommand{\thefigure}{S.\arabic{figure}}
\section{\textit{2.5-dimensional} Forward Model}\label{sec:suppl_2p5_fwd_model}
\begin{figure}[h]
    \centering
    \includegraphics[width=0.9\textwidth]{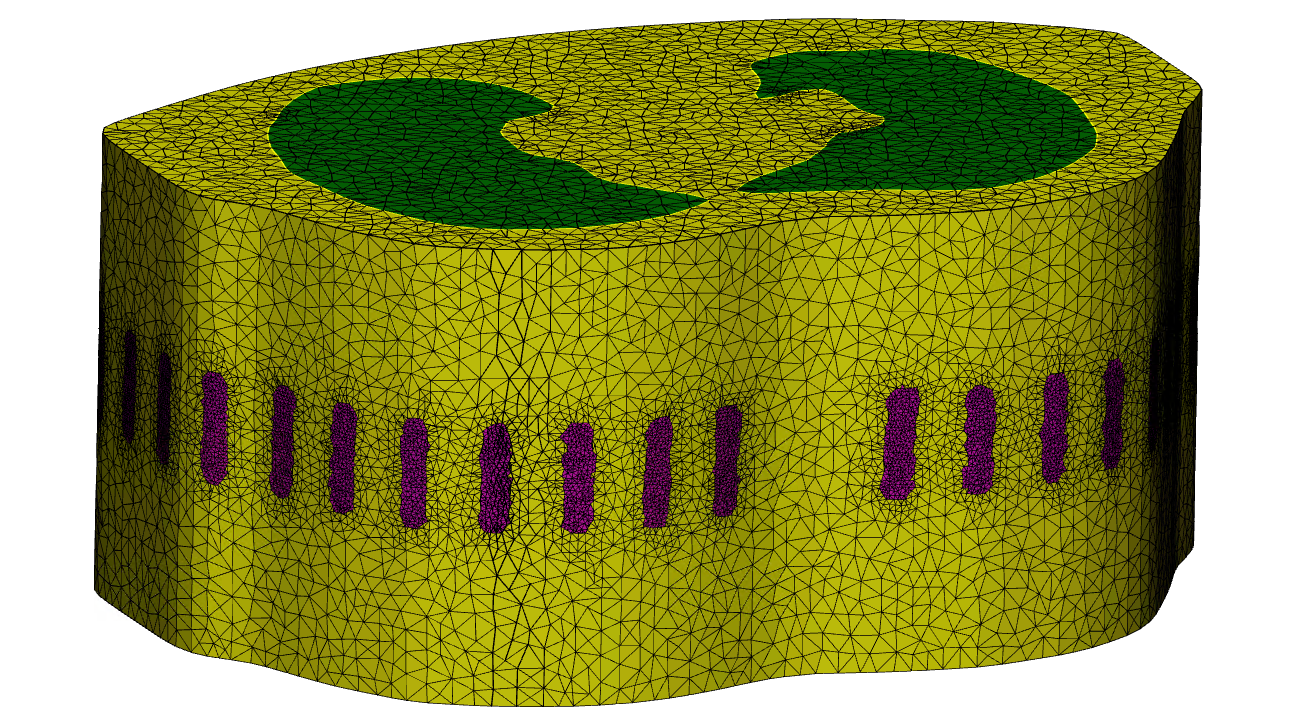}
    \caption{Simplified model geometry based on the extrusion of a two-dimensional slice from the CT scan.}
    \label{fig:2d_plus_geometry}
\end{figure}

The thorax model that is used in the Sections \ref{sec:scenario_setup_results}, \ref{sec:param_study_unif} and \ref{sec:param_study_cust} is based on the CT Scan of an ARDS patient. We obtained the model extruding one transversal plane by $160\,\mathrm{mm}$ along the longitudinal axis, see \Fref{fig:2d_plus_geometry}. 32 electrodes are placed according to the SenTec \textit{SensorBelt} design (SenTec, Landquart, Switzerland), where increased electrode gaps are located in the region of the spine and the sternum. The electrodes have a rectangular shape with the dimensions $\sim 40 \times 10 \,\mathrm{mm}$. To account for the high gradients in the electric field in the vicinity of the electrodes, the global mesh size ($\sim10\,\mathrm{mm}$ edge length) is gradually refined towards the electrode surfaces ($\sim1\,\mathrm{mm}$ edge length). In contrast, to the actual placement of the \textit{SensorBelt} in the clinic, we place the electrodes in an exactly transverse plane to have a well defined reference for the quantitative evaluations in the Sections \ref{sec:param_study_unif} and \ref{sec:param_study_cust}.
\section{3D Patient Model}
\begin{figure}[h]
    \centering
    \includegraphics[width=0.9\textwidth]{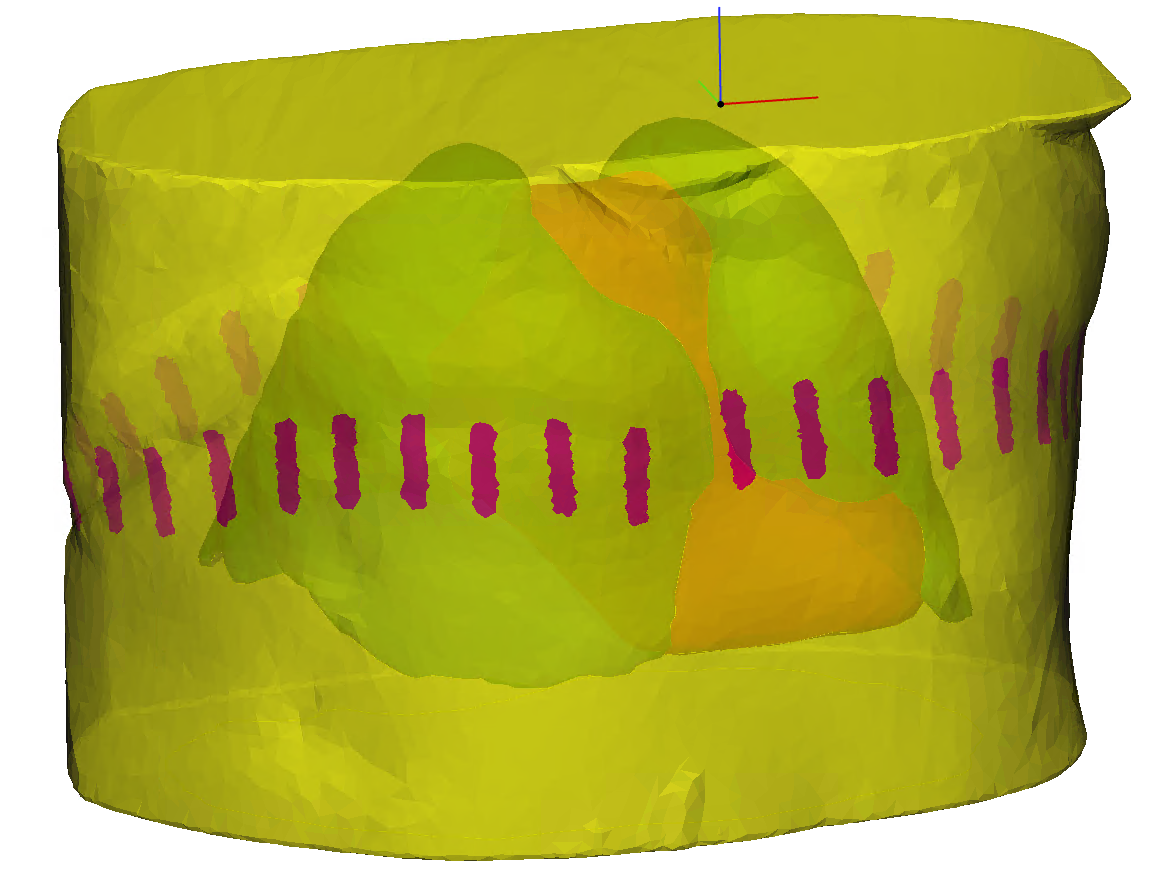}
    \caption{Patient thorax geometry to generate the GREIT training measurements that are used in the showcase reconstruction examples of Section \ref{sec:showcase_e9}.}
    \label{fig:E9_fwd_model}
\end{figure}

In Section \ref{sec:showcase_e9} we showcase how our adapted parametrization affects the EIT images of a mechanically ventilated ARDS patient compared with reconstruction setups from literature. In all three cases that are shown in \Fref{fig:showcase_e9}, we use the same forward model to generate the simulated GREIT training data. \Fref{fig:E9_fwd_model} shows the corresponding geometry, which was obtained by segmenting the CT images of the patient. As the reconstruction of Thürk et al.~(2017) also considers the heart in the background conductivity, we distinguish between four volumes, namely the heart, the left and right lung as well as the thorax. Equivalent to the model setup in Section \ref{sec:suppl_2p5_fwd_model}, we use rectangular electrodes ($\sim 40 \times 10 \,\mathrm{mm}$) which are discretized with a much smaller mesh size ($\sim1\,\mathrm{mm}$) than the rest of the geometry ($\sim10\,\mathrm{mm}$). The considered patient had the EIT \textit{SensorBelt} attached during the recording of the CT image so we were able to use the actual clinical belt position for the placement of the electrodes.

%% file: figures/targets_sketch_annotated_no_boxes.pdf_tex
\begingroup%
  \makeatletter%
  \providecommand\color[2][]{%
    \errmessage{(Inkscape) Color is used for the text in Inkscape, but the package 'color.sty' is not loaded}%
    \renewcommand\color[2][]{}%
  }%
  \providecommand\transparent[1]{%
    \errmessage{(Inkscape) Transparency is used (non-zero) for the text in Inkscape, but the package 'transparent.sty' is not loaded}%
    \renewcommand\transparent[1]{}%
  }%
  \providecommand\rotatebox[2]{#2}%
  \newcommand*\fsize{\dimexpr\f@size pt\relax}%
  \newcommand*\lineheight[1]{\fontsize{\fsize}{#1\fsize}\selectfont}%
  \ifx\svgwidth\undefined%
    \setlength{\unitlength}{722.80467116bp}%
    \ifx\svgscale\undefined%
      \relax%
    \else%
      \setlength{\unitlength}{\unitlength * \real{\svgscale}}%
    \fi%
  \else%
    \setlength{\unitlength}{\svgwidth}%
  \fi%
  \global\let\svgwidth\undefined%
  \global\let\svgscale\undefined%
  \makeatother%
  \begin{picture}(1,0.47327785)%
    \lineheight{1}%
    \setlength\tabcolsep{0pt}%
    \put(0,0){\includegraphics[width=\unitlength,page=1]{targets_sketch_annotated_no_boxes.pdf}}%
    \put(0.67663779,0.08394938){\color[rgb]{0,0,0}\makebox(0,0)[lt]{\lineheight{1.25}\smash{\begin{tabular}[t]{l}$\sigma_{\mathrm{bkg,lung}}$\end{tabular}}}}%
    \put(0.54072026,0.01025209){\color[rgb]{0,0,0}\makebox(0,0)[lt]{\lineheight{1.25}\smash{\begin{tabular}[t]{l}$\sigma_{\mathrm{bkg,thorax}}$\\\end{tabular}}}}%
    \put(0,0){\includegraphics[width=\unitlength,page=2]{targets_sketch_annotated_no_boxes.pdf}}%
    \put(0.1089497,0.00349752){\color[rgb]{0,0,0}\makebox(0,0)[lt]{\lineheight{1.25}\smash{\begin{tabular}[t]{l}$s_{\mathrm{bound}}$\end{tabular}}}}%
    \put(0,0){\includegraphics[width=\unitlength,page=3]{targets_sketch_annotated_no_boxes.pdf}}%
  \end{picture}%
\endgroup%

%% file: figures/sim_scenarios_methodology_annotated.pdf_tex
\begingroup%
  \makeatletter%
  \providecommand\color[2][]{%
    \errmessage{(Inkscape) Color is used for the text in Inkscape, but the package 'color.sty' is not loaded}%
    \renewcommand\color[2][]{}%
  }%
  \providecommand\transparent[1]{%
    \errmessage{(Inkscape) Transparency is used (non-zero) for the text in Inkscape, but the package 'transparent.sty' is not loaded}%
    \renewcommand\transparent[1]{}%
  }%
  \providecommand\rotatebox[2]{#2}%
  \newcommand*\fsize{\dimexpr\f@size pt\relax}%
  \newcommand*\lineheight[1]{\fontsize{\fsize}{#1\fsize}\selectfont}%
  \ifx\svgwidth\undefined%
    \setlength{\unitlength}{675.74998862bp}%
    \ifx\svgscale\undefined%
      \relax%
    \else%
      \setlength{\unitlength}{\unitlength * \real{\svgscale}}%
    \fi%
  \else%
    \setlength{\unitlength}{\svgwidth}%
  \fi%
  \global\let\svgwidth\undefined%
  \global\let\svgscale\undefined%
  \makeatother%
  \begin{picture}(1,1.11098782)%
    \lineheight{1}%
    \setlength\tabcolsep{0pt}%
    \put(0,0){\includegraphics[width=\unitlength,page=1]{sim_scenarios_methodology_annotated.pdf}}%
    \put(0.08411287,1.03607104){\color[rgb]{0,0,0}\makebox(0,0)[lt]{\lineheight{1.25}\smash{\begin{tabular}[t]{l}Ventilation \\scenario\end{tabular}}}}%
    \put(0.08354059,0.79633739){\color[rgb]{0,0,0}\makebox(0,0)[lt]{\lineheight{1.25}\smash{\begin{tabular}[t]{l}Conductivity \\distribution\end{tabular}}}}%
    \put(0.08304346,0.59735226){\color[rgb]{0,0,0}\makebox(0,0)[lt]{\lineheight{1.25}\smash{\begin{tabular}[t]{l}Forward model\end{tabular}}}}%
    \put(0.08411287,0.47907214){\color[rgb]{0,0,0}\makebox(0,0)[lt]{\lineheight{1.25}\smash{\begin{tabular}[t]{l}Voltage difference\end{tabular}}}}%
    \put(0.08304346,0.31497014){\color[rgb]{0,0,0}\makebox(0,0)[lt]{\lineheight{1.25}\smash{\begin{tabular}[t]{l}Reconstruction \\of noisy voltages\end{tabular}}}}%
    \put(0.08411287,0.07603074){\color[rgb]{0,0,0}\makebox(0,0)[lt]{\lineheight{1.25}\smash{\begin{tabular}[t]{l}Average image\end{tabular}}}}%
    \put(0,0){\includegraphics[width=\unitlength,page=2]{sim_scenarios_methodology_annotated.pdf}}%
    \put(1.38649553,0.98611028){\color[rgb]{0,0,0}\makebox(0,0)[lt]{\begin{minipage}{2.00217275\unitlength}\end{minipage}}}%
    \put(0.36722837,0.85122733){\color[rgb]{0,0,0}\makebox(0,0)[lt]{\begin{minipage}{0.27426527\unitlength}\raggedright $\boldsymbol{\sigma}_\mathrm{ref}$\end{minipage}}}%
    \put(0.51893563,0.85122733){\color[rgb]{0,0,0}\makebox(0,0)[lt]{\begin{minipage}{0.27426527\unitlength}\raggedright $\boldsymbol{\sigma}_\mathrm{insp}$\end{minipage}}}%
    \put(0.67771304,0.85122733){\color[rgb]{0,0,0}\makebox(0,0)[lt]{\begin{minipage}{0.27426527\unitlength}\raggedright $\Delta \boldsymbol{\sigma}$\end{minipage}}}%
    \put(0.32491023,0.85122733){\color[rgb]{0,0,0}\makebox(0,0)[lt]{\begin{minipage}{0.27426527\unitlength}\raggedright $-$\end{minipage}}}%
    \put(0.45690643,0.85122733){\color[rgb]{0,0,0}\makebox(0,0)[lt]{\begin{minipage}{0.27426527\unitlength}\raggedright $+$\end{minipage}}}%
    \put(0.62441871,0.85122733){\color[rgb]{0,0,0}\makebox(0,0)[lt]{\begin{minipage}{0.27426527\unitlength}\raggedright $=$\end{minipage}}}%
    \put(0.43947267,0.49214086){\color[rgb]{0,0,0}\makebox(0,0)[lt]{\begin{minipage}{0.27426527\unitlength}\raggedright $\boldsymbol{\mathrm{v}}_\mathrm{ref}$\end{minipage}}}%
    \put(0.59117996,0.49214086){\color[rgb]{0,0,0}\makebox(0,0)[lt]{\begin{minipage}{0.27426527\unitlength}\raggedright $\boldsymbol{\mathrm{v}}_\mathrm{insp}$\end{minipage}}}%
    \put(0.74995734,0.49214086){\color[rgb]{0,0,0}\makebox(0,0)[lt]{\begin{minipage}{0.27426527\unitlength}\raggedright $\Delta \boldsymbol{\mathrm{v}}$\end{minipage}}}%
    \put(0.39715455,0.49214086){\color[rgb]{0,0,0}\makebox(0,0)[lt]{\begin{minipage}{0.27426527\unitlength}\raggedright $-$\end{minipage}}}%
    \put(0.52915073,0.49214086){\color[rgb]{0,0,0}\makebox(0,0)[lt]{\begin{minipage}{0.27426527\unitlength}\raggedright $+$\end{minipage}}}%
    \put(0.69666303,0.49214086){\color[rgb]{0,0,0}\makebox(0,0)[lt]{\begin{minipage}{0.27426527\unitlength}\raggedright $=$\end{minipage}}}%
    \put(0.35447922,0.35183356){\color[rgb]{0,0,0}\makebox(0,0)[lt]{\lineheight{1.25}\smash{\begin{tabular}[t]{l}$\hat{\boldsymbol{\mathrm{x}}}_i = \boldsymbol{\mathrm{R}}\cdot(\Delta \boldsymbol{\mathrm{v}} + \boldsymbol{\mathrm{n}}_i)$\end{tabular}}}}%
    \put(0.69292906,1.05542226){\color[rgb]{0,0,0}\makebox(0,0)[lt]{\lineheight{1.25}\smash{\begin{tabular}[t]{l}non-ventilated (thorax)\end{tabular}}}}%
    \put(0.69292906,1.02212593){\color[rgb]{0,0,0}\makebox(0,0)[lt]{\lineheight{1.25}\smash{\begin{tabular}[t]{l}non-ventilated (lung)\end{tabular}}}}%
    \put(0.69329313,0.98771969){\color[rgb]{0,0,0}\makebox(0,0)[lt]{\lineheight{1.25}\smash{\begin{tabular}[t]{l}ventilated (lung)\end{tabular}}}}%
    \put(0.41247837,0.07396917){\color[rgb]{0,0,0}\makebox(0,0)[lt]{\lineheight{1.25}\smash{\begin{tabular}[t]{l}$\hat{\boldsymbol{\mathrm{x}}}_m = \frac{1}{n_n}\sum\limits_{i=1}^{n_n}\hat{\boldsymbol{\mathrm{x}}}_i$\end{tabular}}}}%
    \put(0.84211079,0.84757901){\color[rgb]{0,0,0}\makebox(0,0)[lt]{\lineheight{1.25}\smash{\begin{tabular}[t]{l}$\sigma_\mathrm{lung, aerated}$\end{tabular}}}}%
    \put(0.84211079,0.81650247){\color[rgb]{0,0,0}\makebox(0,0)[lt]{\lineheight{1.25}\smash{\begin{tabular}[t]{l}$\sigma_\mathrm{lung, non-aerated}$\end{tabular}}}}%
    \put(0.84247493,0.78431599){\color[rgb]{0,0,0}\makebox(0,0)[lt]{\lineheight{1.25}\smash{\begin{tabular}[t]{l}$\sigma_\mathrm{thorax}$\end{tabular}}}}%
    \put(0.84247493,0.75323941){\color[rgb]{0,0,0}\makebox(0,0)[lt]{\lineheight{1.25}\smash{\begin{tabular}[t]{l}$\Delta\sigma_\mathrm{active}$\end{tabular}}}}%
    \put(0.84247493,0.72216283){\color[rgb]{0,0,0}\makebox(0,0)[lt]{\lineheight{1.25}\smash{\begin{tabular}[t]{l}$\Delta\sigma_\mathrm{passive}$\end{tabular}}}}%
    \put(0.63401789,0.35183356){\color[rgb]{0,0,0}\makebox(0,0)[lt]{\lineheight{1.25}\smash{\begin{tabular}[t]{l}$\boldsymbol{\mathrm{n}}_i \in \mathbb{R}^{n_M}$\end{tabular}}}}%
    \put(0.75027805,0.35183356){\color[rgb]{0,0,0}\makebox(0,0)[lt]{\lineheight{1.25}\smash{\begin{tabular}[t]{l}for $i \in \{1,2,\ldots, n_n\}$\end{tabular}}}}%
    \put(0.40638556,0.22696903){\color[rgb]{0,0,0}\makebox(0,0)[lt]{\lineheight{1.25}\smash{\begin{tabular}[t]{l}$\hat{\boldsymbol{\mathrm{x}}}_1$\end{tabular}}}}%
    \put(0.52349264,0.22696903){\color[rgb]{0,0,0}\makebox(0,0)[lt]{\lineheight{1.25}\smash{\begin{tabular}[t]{l}$\hat{\boldsymbol{\mathrm{x}}}_2$\end{tabular}}}}%
    \put(0.64260114,0.22696903){\color[rgb]{0,0,0}\makebox(0,0)[lt]{\lineheight{1.25}\smash{\begin{tabular}[t]{l}$\hat{\boldsymbol{\mathrm{x}}}_3$\end{tabular}}}}%
    \put(0.80459909,0.22696903){\color[rgb]{0,0,0}\makebox(0,0)[lt]{\lineheight{1.25}\smash{\begin{tabular}[t]{l}$\hat{\boldsymbol{\mathrm{x}}}_{n_n}$\end{tabular}}}}%
    \put(0.39694435,0.60234685){\color[rgb]{0,0,0}\makebox(0,0)[lt]{\lineheight{1.25}\smash{\begin{tabular}[t]{l}$\boldsymbol{\mathrm{v}}_\mathrm{ref}=F(\boldsymbol{\sigma}_\mathrm{ref})$\end{tabular}}}}%
    \put(0.69776152,0.60234685){\color[rgb]{0,0,0}\makebox(0,0)[lt]{\lineheight{1.25}\smash{\begin{tabular}[t]{l}$\boldsymbol{\mathrm{v}}_\mathrm{insp}=F(\boldsymbol{\sigma}_\mathrm{insp})$\end{tabular}}}}%
    \put(0.02210234,1.01993839){\color[rgb]{0,0,0}\makebox(0,0)[lt]{\lineheight{1.25}\smash{\begin{tabular}[t]{l}1\end{tabular}}}}%
    \put(0.02231911,0.78035097){\color[rgb]{0,0,0}\makebox(0,0)[lt]{\lineheight{1.25}\smash{\begin{tabular}[t]{l}2\end{tabular}}}}%
    \put(0.02230178,0.59871188){\color[rgb]{0,0,0}\makebox(0,0)[lt]{\lineheight{1.25}\smash{\begin{tabular}[t]{l}3\end{tabular}}}}%
    \put(0.02246363,0.48066082){\color[rgb]{0,0,0}\makebox(0,0)[lt]{\lineheight{1.25}\smash{\begin{tabular}[t]{l}4\end{tabular}}}}%
    \put(0.02229599,0.29901031){\color[rgb]{0,0,0}\makebox(0,0)[lt]{\lineheight{1.25}\smash{\begin{tabular}[t]{l}5\end{tabular}}}}%
    \put(0.02233935,0.07784254){\color[rgb]{0,0,0}\makebox(0,0)[lt]{\lineheight{1.25}\smash{\begin{tabular}[t]{l}6\end{tabular}}}}%
  \end{picture}%
\endgroup%

%% file: figures/sim_scenarios_with_ref_annotated.pdf_tex
\begingroup%
  \makeatletter%
  \providecommand\color[2][]{%
    \errmessage{(Inkscape) Color is used for the text in Inkscape, but the package 'color.sty' is not loaded}%
    \renewcommand\color[2][]{}%
  }%
  \providecommand\transparent[1]{%
    \errmessage{(Inkscape) Transparency is used (non-zero) for the text in Inkscape, but the package 'transparent.sty' is not loaded}%
    \renewcommand\transparent[1]{}%
  }%
  \providecommand\rotatebox[2]{#2}%
  \newcommand*\fsize{\dimexpr\f@size pt\relax}%
  \newcommand*\lineheight[1]{\fontsize{\fsize}{#1\fsize}\selectfont}%
  \ifx\svgwidth\undefined%
    \setlength{\unitlength}{693.48382413bp}%
    \ifx\svgscale\undefined%
      \relax%
    \else%
      \setlength{\unitlength}{\unitlength * \real{\svgscale}}%
    \fi%
  \else%
    \setlength{\unitlength}{\svgwidth}%
  \fi%
  \global\let\svgwidth\undefined%
  \global\let\svgscale\undefined%
  \makeatother%
  \begin{picture}(1,0.17191653)%
    \lineheight{1}%
    \setlength\tabcolsep{0pt}%
    \put(0,0){\includegraphics[width=\unitlength,page=1]{sim_scenarios_with_ref_annotated.pdf}}%
    \put(0.19992585,0.15658159){\color[rgb]{0,0,0}\makebox(0,0)[lt]{\lineheight{1.25}\smash{\begin{tabular}[t]{l}Scenario 1\end{tabular}}}}%
    \put(0.36554146,0.15658159){\color[rgb]{0,0,0}\makebox(0,0)[lt]{\lineheight{1.25}\smash{\begin{tabular}[t]{l}Scenario 2\end{tabular}}}}%
    \put(0.53102191,0.15658159){\color[rgb]{0,0,0}\makebox(0,0)[lt]{\lineheight{1.25}\smash{\begin{tabular}[t]{l}Scenario 3\end{tabular}}}}%
    \put(0.69684873,0.15658159){\color[rgb]{0,0,0}\makebox(0,0)[lt]{\lineheight{1.25}\smash{\begin{tabular}[t]{l}Scenario 4\end{tabular}}}}%
    \put(0.86308958,0.15658159){\color[rgb]{0,0,0}\makebox(0,0)[lt]{\lineheight{1.25}\smash{\begin{tabular}[t]{l}Scenario 5\end{tabular}}}}%
    \put(0,0){\includegraphics[width=\unitlength,page=2]{sim_scenarios_with_ref_annotated.pdf}}%
    \put(0.05899045,0.00119978){\color[rgb]{0,0,0}\makebox(0,0)[lt]{\lineheight{1.25}\smash{\begin{tabular}[t]{l}Active Lung:\end{tabular}}}}%
    \put(0,0){\includegraphics[width=\unitlength,page=3]{sim_scenarios_with_ref_annotated.pdf}}%
    \put(0.40719451,0.00119978){\color[rgb]{0,0,0}\makebox(0,0)[lt]{\lineheight{1.25}\smash{\begin{tabular}[t]{l}Passive Lung: \end{tabular}}}}%
    \put(0,0){\includegraphics[width=\unitlength,page=4]{sim_scenarios_with_ref_annotated.pdf}}%
    \put(0.79848067,0.00119978){\color[rgb]{0,0,0}\makebox(0,0)[lt]{\lineheight{1.25}\smash{\begin{tabular}[t]{l}Thorax:\end{tabular}}}}%
    \put(0.03390468,0.15658159){\color[rgb]{0,0,0}\makebox(0,0)[lt]{\lineheight{1.25}\smash{\begin{tabular}[t]{l}Reference\end{tabular}}}}%
    \put(0.55674984,0.00336278){\color[rgb]{0,0,0}\makebox(0,0)[lt]{\lineheight{1.25}\smash{\begin{tabular}[t]{l}$\sigma_{\mathrm{lung,\,non-aerated}}$\end{tabular}}}}%
    \put(0.20287553,0.00336278){\color[rgb]{0,0,0}\makebox(0,0)[lt]{\lineheight{1.25}\smash{\begin{tabular}[t]{l}$\sigma_{\mathrm{lung,\,aerated}}$\end{tabular}}}}%
    \put(0.89070451,0.00336278){\color[rgb]{0,0,0}\makebox(0,0)[lt]{\lineheight{1.25}\smash{\begin{tabular}[t]{l}$\sigma_{\mathrm{thorax}}$\end{tabular}}}}%
  \end{picture}%
\endgroup%

%% file: figures/noise_samples_uniform.pdf_tex
\begingroup%
  \makeatletter%
  \providecommand\color[2][]{%
    \errmessage{(Inkscape) Color is used for the text in Inkscape, but the package 'color.sty' is not loaded}%
    \renewcommand\color[2][]{}%
  }%
  \providecommand\transparent[1]{%
    \errmessage{(Inkscape) Transparency is used (non-zero) for the text in Inkscape, but the package 'transparent.sty' is not loaded}%
    \renewcommand\transparent[1]{}%
  }%
  \providecommand\rotatebox[2]{#2}%
  \newcommand*\fsize{\dimexpr\f@size pt\relax}%
  \newcommand*\lineheight[1]{\fontsize{\fsize}{#1\fsize}\selectfont}%
  \ifx\svgwidth\undefined%
    \setlength{\unitlength}{656.25021921bp}%
    \ifx\svgscale\undefined%
      \relax%
    \else%
      \setlength{\unitlength}{\unitlength * \real{\svgscale}}%
    \fi%
  \else%
    \setlength{\unitlength}{\svgwidth}%
  \fi%
  \global\let\svgwidth\undefined%
  \global\let\svgscale\undefined%
  \makeatother%
  \begin{picture}(1,0.36552411)%
    \lineheight{1}%
    \setlength\tabcolsep{0pt}%
    \put(0,0){\includegraphics[width=\unitlength,page=1]{noise_samples_uniform.pdf}}%
    \put(0.50352598,0.3562622){\color[rgb]{0,0,0}\makebox(0,0)[lt]{\lineheight{1.25}\smash{\begin{tabular}[t]{l}Noise Samples\end{tabular}}}}%
    \put(0.37491197,0.24640489){\color[rgb]{0,0,0}\makebox(0,0)[lt]{\lineheight{1.25}\smash{\begin{tabular}[t]{l}$\mathrm{SNR_{im}} = 0.913\, \mathrm{dB}$\end{tabular}}}}%
    \put(0,0){\includegraphics[width=\unitlength,page=2]{noise_samples_uniform.pdf}}%
    \put(0.37491197,0.13019065){\color[rgb]{0,0,0}\makebox(0,0)[lt]{\lineheight{1.25}\smash{\begin{tabular}[t]{l}$\mathrm{SNR_{im}} = 7.969\, \mathrm{dB}$\end{tabular}}}}%
    \put(0,0){\includegraphics[width=\unitlength,page=3]{noise_samples_uniform.pdf}}%
    \put(0.37491197,0.01397635){\color[rgb]{0,0,0}\makebox(0,0)[lt]{\lineheight{1.25}\smash{\begin{tabular}[t]{l}$\mathrm{SNR_{im}} = 5.761 \,\mathrm{dB}$\end{tabular}}}}%
    \put(0.19900811,0.3562622){\color[rgb]{0,0,0}\makebox(0,0)[lt]{\lineheight{1.25}\smash{\begin{tabular}[t]{l}Vent. Scenario\end{tabular}}}}%
    \put(0.06650496,0.30119484){\color[rgb]{0,0,0}\makebox(0,0)[lt]{\lineheight{1.25}\smash{\begin{tabular}[t]{l}$\gamma_\sigma = 0.2$ \end{tabular}}}}%
    \put(0.06650496,0.27102853){\color[rgb]{0,0,0}\makebox(0,0)[lt]{\lineheight{1.25}\smash{\begin{tabular}[t]{l}$\mathrm{NF} = 0.9$\end{tabular}}}}%
    \put(0.06650496,0.18498067){\color[rgb]{0,0,0}\makebox(0,0)[lt]{\lineheight{1.25}\smash{\begin{tabular}[t]{l}$\gamma_\sigma = 0.2$ \end{tabular}}}}%
    \put(0.06650496,0.15481436){\color[rgb]{0,0,0}\makebox(0,0)[lt]{\lineheight{1.25}\smash{\begin{tabular}[t]{l}$\mathrm{NF} = 0.3$\end{tabular}}}}%
    \put(0.06650495,0.06876647){\color[rgb]{0,0,0}\makebox(0,0)[lt]{\lineheight{1.25}\smash{\begin{tabular}[t]{l}$\gamma_\sigma = 1.0$ \end{tabular}}}}%
    \put(0.06650495,0.03860016){\color[rgb]{0,0,0}\makebox(0,0)[lt]{\lineheight{1.25}\smash{\begin{tabular}[t]{l}$\mathrm{NF} = 0.9$\end{tabular}}}}%
    \put(0.8284118,0.3562622){\color[rgb]{0,0,0}\makebox(0,0)[lt]{\lineheight{1.25}\smash{\begin{tabular}[t]{l}Averaged Image\end{tabular}}}}%
    \put(0.01005342,0.28592719){\color[rgb]{0,0,0}\makebox(0,0)[lt]{\lineheight{1.25}\smash{\begin{tabular}[t]{l}(a)\end{tabular}}}}%
    \put(0.01005342,0.16970702){\color[rgb]{0,0,0}\makebox(0,0)[lt]{\lineheight{1.25}\smash{\begin{tabular}[t]{l}(b)\end{tabular}}}}%
    \put(0.01005342,0.05349891){\color[rgb]{0,0,0}\makebox(0,0)[lt]{\lineheight{1.25}\smash{\begin{tabular}[t]{l}(c)\end{tabular}}}}%
    \put(0,0){\includegraphics[width=\unitlength,page=4]{noise_samples_uniform.pdf}}%
  \end{picture}%
\endgroup%

%% file: figures/param_space_uniform_target.pdf_tex
\begingroup%
  \makeatletter%
  \providecommand\color[2][]{%
    \errmessage{(Inkscape) Color is used for the text in Inkscape, but the package 'color.sty' is not loaded}%
    \renewcommand\color[2][]{}%
  }%
  \providecommand\transparent[1]{%
    \errmessage{(Inkscape) Transparency is used (non-zero) for the text in Inkscape, but the package 'transparent.sty' is not loaded}%
    \renewcommand\transparent[1]{}%
  }%
  \providecommand\rotatebox[2]{#2}%
  \newcommand*\fsize{\dimexpr\f@size pt\relax}%
  \newcommand*\lineheight[1]{\fontsize{\fsize}{#1\fsize}\selectfont}%
  \ifx\svgwidth\undefined%
    \setlength{\unitlength}{575.9999856bp}%
    \ifx\svgscale\undefined%
      \relax%
    \else%
      \setlength{\unitlength}{\unitlength * \real{\svgscale}}%
    \fi%
  \else%
    \setlength{\unitlength}{\svgwidth}%
  \fi%
  \global\let\svgwidth\undefined%
  \global\let\svgscale\undefined%
  \makeatother%
  \begin{picture}(1,0.5)%
    \lineheight{1}%
    \setlength\tabcolsep{0pt}%
    \put(0,0){\includegraphics[width=\unitlength,page=1]{param_space_uniform_target.pdf}}%
    \put(0.08974764,0.04751085){\color[rgb]{0,0,0}\makebox(0,0)[lt]{\lineheight{1.25}\smash{\begin{tabular}[t]{l}0.2\end{tabular}}}}%
    \put(0,0){\includegraphics[width=\unitlength,page=2]{param_space_uniform_target.pdf}}%
    \put(0.1718762,0.04751085){\color[rgb]{0,0,0}\makebox(0,0)[lt]{\lineheight{1.25}\smash{\begin{tabular}[t]{l}0.4\end{tabular}}}}%
    \put(0,0){\includegraphics[width=\unitlength,page=3]{param_space_uniform_target.pdf}}%
    \put(0.25400476,0.04751085){\color[rgb]{0,0,0}\makebox(0,0)[lt]{\lineheight{1.25}\smash{\begin{tabular}[t]{l}0.6\end{tabular}}}}%
    \put(0,0){\includegraphics[width=\unitlength,page=4]{param_space_uniform_target.pdf}}%
    \put(0.33613332,0.04751085){\color[rgb]{0,0,0}\makebox(0,0)[lt]{\lineheight{1.25}\smash{\begin{tabular}[t]{l}0.8\end{tabular}}}}%
    \put(0.22111036,0.02377496){\color[rgb]{0,0,0}\makebox(0,0)[lt]{\lineheight{1.25}\smash{\begin{tabular}[t]{l}$\mathrm{NF}$ \end{tabular}}}}%
    \put(0,0){\includegraphics[width=\unitlength,page=5]{param_space_uniform_target.pdf}}%
    \put(0.04325521,0.10836108){\color[rgb]{0,0,0}\makebox(0,0)[lt]{\lineheight{1.25}\smash{\begin{tabular}[t]{l}0.5\end{tabular}}}}%
    \put(0,0){\includegraphics[width=\unitlength,page=6]{param_space_uniform_target.pdf}}%
    \put(0.04325521,0.17853717){\color[rgb]{0,0,0}\makebox(0,0)[lt]{\lineheight{1.25}\smash{\begin{tabular}[t]{l}1.0\end{tabular}}}}%
    \put(0,0){\includegraphics[width=\unitlength,page=7]{param_space_uniform_target.pdf}}%
    \put(0.04325521,0.24871326){\color[rgb]{0,0,0}\makebox(0,0)[lt]{\lineheight{1.25}\smash{\begin{tabular}[t]{l}1.5\end{tabular}}}}%
    \put(0,0){\includegraphics[width=\unitlength,page=8]{param_space_uniform_target.pdf}}%
    \put(0.04325521,0.31888936){\color[rgb]{0,0,0}\makebox(0,0)[lt]{\lineheight{1.25}\smash{\begin{tabular}[t]{l}2.0\end{tabular}}}}%
    \put(0,0){\includegraphics[width=\unitlength,page=9]{param_space_uniform_target.pdf}}%
    \put(0.04325521,0.38906545){\color[rgb]{0,0,0}\makebox(0,0)[lt]{\lineheight{1.25}\smash{\begin{tabular}[t]{l}2.5\end{tabular}}}}%
    \put(0,0){\includegraphics[width=\unitlength,page=10]{param_space_uniform_target.pdf}}%
    \put(0.04325521,0.45924153){\color[rgb]{0,0,0}\makebox(0,0)[lt]{\lineheight{1.25}\smash{\begin{tabular}[t]{l}3.0\end{tabular}}}}%
    \put(0.03221463,0.27155165){\color[rgb]{0,0,0}\rotatebox{90}{\makebox(0,0)[lt]{\lineheight{1.25}\smash{\begin{tabular}[t]{l}$\gamma_\sigma$ \end{tabular}}}}}%
    \put(0,0){\includegraphics[width=\unitlength,page=11]{param_space_uniform_target.pdf}}%
    \put(0.13587202,0.47625){\color[rgb]{0,0,0}\makebox(0,0)[lt]{\lineheight{1.25}\smash{\begin{tabular}[t]{l}$\mathrm{mean}(\mathrm{SNR_{im}}) (\mathrm{dB})$\end{tabular}}}}%
    \put(0,0){\includegraphics[width=\unitlength,page=12]{param_space_uniform_target.pdf}}%
    \put(0.57490389,0.04751085){\color[rgb]{0,0,0}\makebox(0,0)[lt]{\lineheight{1.25}\smash{\begin{tabular}[t]{l}0.2\end{tabular}}}}%
    \put(0,0){\includegraphics[width=\unitlength,page=13]{param_space_uniform_target.pdf}}%
    \put(0.65703245,0.04751085){\color[rgb]{0,0,0}\makebox(0,0)[lt]{\lineheight{1.25}\smash{\begin{tabular}[t]{l}0.4\end{tabular}}}}%
    \put(0,0){\includegraphics[width=\unitlength,page=14]{param_space_uniform_target.pdf}}%
    \put(0.73916101,0.04751085){\color[rgb]{0,0,0}\makebox(0,0)[lt]{\lineheight{1.25}\smash{\begin{tabular}[t]{l}0.6\end{tabular}}}}%
    \put(0,0){\includegraphics[width=\unitlength,page=15]{param_space_uniform_target.pdf}}%
    \put(0.82128957,0.04751085){\color[rgb]{0,0,0}\makebox(0,0)[lt]{\lineheight{1.25}\smash{\begin{tabular}[t]{l}0.8\end{tabular}}}}%
    \put(0.70626661,0.02377496){\color[rgb]{0,0,0}\makebox(0,0)[lt]{\lineheight{1.25}\smash{\begin{tabular}[t]{l}$\mathrm{NF}$ \end{tabular}}}}%
    \put(0,0){\includegraphics[width=\unitlength,page=16]{param_space_uniform_target.pdf}}%
    \put(0.52841146,0.10836108){\color[rgb]{0,0,0}\makebox(0,0)[lt]{\lineheight{1.25}\smash{\begin{tabular}[t]{l}0.5\end{tabular}}}}%
    \put(0,0){\includegraphics[width=\unitlength,page=17]{param_space_uniform_target.pdf}}%
    \put(0.52841146,0.17853717){\color[rgb]{0,0,0}\makebox(0,0)[lt]{\lineheight{1.25}\smash{\begin{tabular}[t]{l}1.0\end{tabular}}}}%
    \put(0,0){\includegraphics[width=\unitlength,page=18]{param_space_uniform_target.pdf}}%
    \put(0.52841146,0.24871326){\color[rgb]{0,0,0}\makebox(0,0)[lt]{\lineheight{1.25}\smash{\begin{tabular}[t]{l}1.5\end{tabular}}}}%
    \put(0,0){\includegraphics[width=\unitlength,page=19]{param_space_uniform_target.pdf}}%
    \put(0.52841146,0.31888936){\color[rgb]{0,0,0}\makebox(0,0)[lt]{\lineheight{1.25}\smash{\begin{tabular}[t]{l}2.0\end{tabular}}}}%
    \put(0,0){\includegraphics[width=\unitlength,page=20]{param_space_uniform_target.pdf}}%
    \put(0.52841146,0.38906545){\color[rgb]{0,0,0}\makebox(0,0)[lt]{\lineheight{1.25}\smash{\begin{tabular}[t]{l}2.5\end{tabular}}}}%
    \put(0,0){\includegraphics[width=\unitlength,page=21]{param_space_uniform_target.pdf}}%
    \put(0.52841146,0.45924153){\color[rgb]{0,0,0}\makebox(0,0)[lt]{\lineheight{1.25}\smash{\begin{tabular}[t]{l}3.0\end{tabular}}}}%
    \put(0.51737088,0.27155165){\color[rgb]{0,0,0}\rotatebox{90}{\makebox(0,0)[lt]{\lineheight{1.25}\smash{\begin{tabular}[t]{l}$\gamma_\sigma$ \end{tabular}}}}}%
    \put(0,0){\includegraphics[width=\unitlength,page=22]{param_space_uniform_target.pdf}}%
    \put(0.65385097,0.47625){\color[rgb]{0,0,0}\makebox(0,0)[lt]{\lineheight{1.25}\smash{\begin{tabular}[t]{l}$\mathrm{mean} (r_{\Delta\boldsymbol{\sigma}})$\end{tabular}}}}%
    \put(0,0){\includegraphics[width=\unitlength,page=23]{param_space_uniform_target.pdf}}%
    \put(0.44205608,0.06625543){\color[rgb]{0,0,0}\makebox(0,0)[lt]{\lineheight{1.25}\smash{\begin{tabular}[t]{l}1.2\end{tabular}}}}%
    \put(0,0){\includegraphics[width=\unitlength,page=24]{param_space_uniform_target.pdf}}%
    \put(0.44205608,0.11389011){\color[rgb]{0,0,0}\makebox(0,0)[lt]{\lineheight{1.25}\smash{\begin{tabular}[t]{l}2.8\end{tabular}}}}%
    \put(0,0){\includegraphics[width=\unitlength,page=25]{param_space_uniform_target.pdf}}%
    \put(0.44205608,0.16152479){\color[rgb]{0,0,0}\makebox(0,0)[lt]{\lineheight{1.25}\smash{\begin{tabular}[t]{l}4.4\end{tabular}}}}%
    \put(0,0){\includegraphics[width=\unitlength,page=26]{param_space_uniform_target.pdf}}%
    \put(0.44205608,0.20915946){\color[rgb]{0,0,0}\makebox(0,0)[lt]{\lineheight{1.25}\smash{\begin{tabular}[t]{l}6.0\end{tabular}}}}%
    \put(0,0){\includegraphics[width=\unitlength,page=27]{param_space_uniform_target.pdf}}%
    \put(0.44205608,0.25679415){\color[rgb]{0,0,0}\makebox(0,0)[lt]{\lineheight{1.25}\smash{\begin{tabular}[t]{l}7.6\end{tabular}}}}%
    \put(0,0){\includegraphics[width=\unitlength,page=28]{param_space_uniform_target.pdf}}%
    \put(0.44205608,0.30442882){\color[rgb]{0,0,0}\makebox(0,0)[lt]{\lineheight{1.25}\smash{\begin{tabular}[t]{l}9.2\end{tabular}}}}%
    \put(0,0){\includegraphics[width=\unitlength,page=29]{param_space_uniform_target.pdf}}%
    \put(0.44205608,0.35206351){\color[rgb]{0,0,0}\makebox(0,0)[lt]{\lineheight{1.25}\smash{\begin{tabular}[t]{l}10.8\end{tabular}}}}%
    \put(0,0){\includegraphics[width=\unitlength,page=30]{param_space_uniform_target.pdf}}%
    \put(0.44205608,0.39969819){\color[rgb]{0,0,0}\makebox(0,0)[lt]{\lineheight{1.25}\smash{\begin{tabular}[t]{l}12.4\end{tabular}}}}%
    \put(0,0){\includegraphics[width=\unitlength,page=31]{param_space_uniform_target.pdf}}%
    \put(0.44205608,0.44733286){\color[rgb]{0,0,0}\makebox(0,0)[lt]{\lineheight{1.25}\smash{\begin{tabular}[t]{l}14.0\end{tabular}}}}%
    \put(0,0){\includegraphics[width=\unitlength,page=32]{param_space_uniform_target.pdf}}%
    \put(0.92721233,0.06625543){\color[rgb]{0,0,0}\makebox(0,0)[lt]{\lineheight{1.25}\smash{\begin{tabular}[t]{l}0.375\end{tabular}}}}%
    \put(0,0){\includegraphics[width=\unitlength,page=33]{param_space_uniform_target.pdf}}%
    \put(0.92721233,0.10690916){\color[rgb]{0,0,0}\makebox(0,0)[lt]{\lineheight{1.25}\smash{\begin{tabular}[t]{l}0.420\end{tabular}}}}%
    \put(0,0){\includegraphics[width=\unitlength,page=34]{param_space_uniform_target.pdf}}%
    \put(0.92721233,0.1475629){\color[rgb]{0,0,0}\makebox(0,0)[lt]{\lineheight{1.25}\smash{\begin{tabular}[t]{l}0.465\end{tabular}}}}%
    \put(0,0){\includegraphics[width=\unitlength,page=35]{param_space_uniform_target.pdf}}%
    \put(0.92721233,0.18821663){\color[rgb]{0,0,0}\makebox(0,0)[lt]{\lineheight{1.25}\smash{\begin{tabular}[t]{l}0.510\end{tabular}}}}%
    \put(0,0){\includegraphics[width=\unitlength,page=36]{param_space_uniform_target.pdf}}%
    \put(0.92721233,0.22887036){\color[rgb]{0,0,0}\makebox(0,0)[lt]{\lineheight{1.25}\smash{\begin{tabular}[t]{l}0.555\end{tabular}}}}%
    \put(0,0){\includegraphics[width=\unitlength,page=37]{param_space_uniform_target.pdf}}%
    \put(0.92721233,0.2695241){\color[rgb]{0,0,0}\makebox(0,0)[lt]{\lineheight{1.25}\smash{\begin{tabular}[t]{l}0.600\end{tabular}}}}%
    \put(0,0){\includegraphics[width=\unitlength,page=38]{param_space_uniform_target.pdf}}%
    \put(0.92721233,0.31017785){\color[rgb]{0,0,0}\makebox(0,0)[lt]{\lineheight{1.25}\smash{\begin{tabular}[t]{l}0.645\end{tabular}}}}%
    \put(0,0){\includegraphics[width=\unitlength,page=39]{param_space_uniform_target.pdf}}%
    \put(0.92721233,0.35083158){\color[rgb]{0,0,0}\makebox(0,0)[lt]{\lineheight{1.25}\smash{\begin{tabular}[t]{l}0.690\end{tabular}}}}%
    \put(0,0){\includegraphics[width=\unitlength,page=40]{param_space_uniform_target.pdf}}%
    \put(0.92721233,0.39148531){\color[rgb]{0,0,0}\makebox(0,0)[lt]{\lineheight{1.25}\smash{\begin{tabular}[t]{l}0.735\end{tabular}}}}%
    \put(0,0){\includegraphics[width=\unitlength,page=41]{param_space_uniform_target.pdf}}%
    \put(0.92721233,0.43213905){\color[rgb]{0,0,0}\makebox(0,0)[lt]{\lineheight{1.25}\smash{\begin{tabular}[t]{l}0.780\end{tabular}}}}%
    \put(0,0){\includegraphics[width=\unitlength,page=42]{param_space_uniform_target.pdf}}%
  \end{picture}%
\endgroup%

%% file: figures/noise_samples_custom.pdf_tex
\begingroup%
  \makeatletter%
  \providecommand\color[2][]{%
    \errmessage{(Inkscape) Color is used for the text in Inkscape, but the package 'color.sty' is not loaded}%
    \renewcommand\color[2][]{}%
  }%
  \providecommand\transparent[1]{%
    \errmessage{(Inkscape) Transparency is used (non-zero) for the text in Inkscape, but the package 'transparent.sty' is not loaded}%
    \renewcommand\transparent[1]{}%
  }%
  \providecommand\rotatebox[2]{#2}%
  \newcommand*\fsize{\dimexpr\f@size pt\relax}%
  \newcommand*\lineheight[1]{\fontsize{\fsize}{#1\fsize}\selectfont}%
  \ifx\svgwidth\undefined%
    \setlength{\unitlength}{656.25021921bp}%
    \ifx\svgscale\undefined%
      \relax%
    \else%
      \setlength{\unitlength}{\unitlength * \real{\svgscale}}%
    \fi%
  \else%
    \setlength{\unitlength}{\svgwidth}%
  \fi%
  \global\let\svgwidth\undefined%
  \global\let\svgscale\undefined%
  \makeatother%
  \begin{picture}(1,0.36552411)%
    \lineheight{1}%
    \setlength\tabcolsep{0pt}%
    \put(0,0){\includegraphics[width=\unitlength,page=1]{noise_samples_custom.pdf}}%
    \put(0.50352598,0.3562622){\color[rgb]{0,0,0}\makebox(0,0)[lt]{\lineheight{1.25}\smash{\begin{tabular}[t]{l}Noise Samples\end{tabular}}}}%
    \put(0.37491197,0.24640489){\color[rgb]{0,0,0}\makebox(0,0)[lt]{\lineheight{1.25}\smash{\begin{tabular}[t]{l}$\mathrm{SNR_{im}} = 7.719\, \mathrm{dB}$\end{tabular}}}}%
    \put(0,0){\includegraphics[width=\unitlength,page=2]{noise_samples_custom.pdf}}%
    \put(0.37491197,0.13019065){\color[rgb]{0,0,0}\makebox(0,0)[lt]{\lineheight{1.25}\smash{\begin{tabular}[t]{l}$\mathrm{SNR_{im}} = 8.737\, \mathrm{dB}$\end{tabular}}}}%
    \put(0,0){\includegraphics[width=\unitlength,page=3]{noise_samples_custom.pdf}}%
    \put(0.37491197,0.01397635){\color[rgb]{0,0,0}\makebox(0,0)[lt]{\lineheight{1.25}\smash{\begin{tabular}[t]{l}$\mathrm{SNR_{im}} = 9.150 \,\mathrm{dB}$\end{tabular}}}}%
    \put(0.19900811,0.3562622){\color[rgb]{0,0,0}\makebox(0,0)[lt]{\lineheight{1.25}\smash{\begin{tabular}[t]{l}Vent. Scenario\end{tabular}}}}%
    \put(0.06650496,0.30119484){\color[rgb]{0,0,0}\makebox(0,0)[lt]{\lineheight{1.25}\smash{\begin{tabular}[t]{l}$\gamma_\rho = 1.0$ \end{tabular}}}}%
    \put(0.06650496,0.27102853){\color[rgb]{0,0,0}\makebox(0,0)[lt]{\lineheight{1.25}\smash{\begin{tabular}[t]{l}$R_w = 0.2$\end{tabular}}}}%
    \put(0.06650496,0.18498067){\color[rgb]{0,0,0}\makebox(0,0)[lt]{\lineheight{1.25}\smash{\begin{tabular}[t]{l}$\gamma_\rho = 3.5$ \end{tabular}}}}%
    \put(0.06650496,0.15481436){\color[rgb]{0,0,0}\makebox(0,0)[lt]{\lineheight{1.25}\smash{\begin{tabular}[t]{l}$R_w = 0.2$\end{tabular}}}}%
    \put(0.06650495,0.06876647){\color[rgb]{0,0,0}\makebox(0,0)[lt]{\lineheight{1.25}\smash{\begin{tabular}[t]{l}$\gamma_\rho = 3.5$ \end{tabular}}}}%
    \put(0.06650495,0.03860016){\color[rgb]{0,0,0}\makebox(0,0)[lt]{\lineheight{1.25}\smash{\begin{tabular}[t]{l}$R_w = 0.1$\end{tabular}}}}%
    \put(0.8284118,0.3562622){\color[rgb]{0,0,0}\makebox(0,0)[lt]{\lineheight{1.25}\smash{\begin{tabular}[t]{l}Averaged Image\end{tabular}}}}%
    \put(0.01005342,0.28592719){\color[rgb]{0,0,0}\makebox(0,0)[lt]{\lineheight{1.25}\smash{\begin{tabular}[t]{l}(a)\end{tabular}}}}%
    \put(0.01005342,0.16970702){\color[rgb]{0,0,0}\makebox(0,0)[lt]{\lineheight{1.25}\smash{\begin{tabular}[t]{l}(b)\end{tabular}}}}%
    \put(0.01005342,0.05349891){\color[rgb]{0,0,0}\makebox(0,0)[lt]{\lineheight{1.25}\smash{\begin{tabular}[t]{l}(c)\end{tabular}}}}%
    \put(0,0){\includegraphics[width=\unitlength,page=4]{noise_samples_custom.pdf}}%
  \end{picture}%
\endgroup%

%% file: figures/param_space_custom_targets.pdf_tex
\begingroup%
  \makeatletter%
  \providecommand\color[2][]{%
    \errmessage{(Inkscape) Color is used for the text in Inkscape, but the package 'color.sty' is not loaded}%
    \renewcommand\color[2][]{}%
  }%
  \providecommand\transparent[1]{%
    \errmessage{(Inkscape) Transparency is used (non-zero) for the text in Inkscape, but the package 'transparent.sty' is not loaded}%
    \renewcommand\transparent[1]{}%
  }%
  \providecommand\rotatebox[2]{#2}%
  \newcommand*\fsize{\dimexpr\f@size pt\relax}%
  \newcommand*\lineheight[1]{\fontsize{\fsize}{#1\fsize}\selectfont}%
  \ifx\svgwidth\undefined%
    \setlength{\unitlength}{575.9999856bp}%
    \ifx\svgscale\undefined%
      \relax%
    \else%
      \setlength{\unitlength}{\unitlength * \real{\svgscale}}%
    \fi%
  \else%
    \setlength{\unitlength}{\svgwidth}%
  \fi%
  \global\let\svgwidth\undefined%
  \global\let\svgscale\undefined%
  \makeatother%
  \begin{picture}(1,0.5)%
    \lineheight{1}%
    \setlength\tabcolsep{0pt}%
    \put(0,0){\includegraphics[width=\unitlength,page=1]{param_space_custom_targets.pdf}}%
    \put(0.08030924,0.04768663){\color[rgb]{0,0,0}\makebox(0,0)[lt]{\lineheight{1.25}\smash{\begin{tabular}[t]{l}1.0\end{tabular}}}}%
    \put(0,0){\includegraphics[width=\unitlength,page=2]{param_space_custom_targets.pdf}}%
    \put(0.14010566,0.04768663){\color[rgb]{0,0,0}\makebox(0,0)[lt]{\lineheight{1.25}\smash{\begin{tabular}[t]{l}1.5\end{tabular}}}}%
    \put(0,0){\includegraphics[width=\unitlength,page=3]{param_space_custom_targets.pdf}}%
    \put(0.19990208,0.04768663){\color[rgb]{0,0,0}\makebox(0,0)[lt]{\lineheight{1.25}\smash{\begin{tabular}[t]{l}2.0\end{tabular}}}}%
    \put(0,0){\includegraphics[width=\unitlength,page=4]{param_space_custom_targets.pdf}}%
    \put(0.25969851,0.04768663){\color[rgb]{0,0,0}\makebox(0,0)[lt]{\lineheight{1.25}\smash{\begin{tabular}[t]{l}2.5\end{tabular}}}}%
    \put(0,0){\includegraphics[width=\unitlength,page=5]{param_space_custom_targets.pdf}}%
    \put(0.31949493,0.04768663){\color[rgb]{0,0,0}\makebox(0,0)[lt]{\lineheight{1.25}\smash{\begin{tabular}[t]{l}3.0\end{tabular}}}}%
    \put(0,0){\includegraphics[width=\unitlength,page=6]{param_space_custom_targets.pdf}}%
    \put(0.37929135,0.04768663){\color[rgb]{0,0,0}\makebox(0,0)[lt]{\lineheight{1.25}\smash{\begin{tabular}[t]{l}3.5\end{tabular}}}}%
    \put(0.21718198,0.02395074){\color[rgb]{0,0,0}\makebox(0,0)[lt]{\lineheight{1.25}\smash{\begin{tabular}[t]{l}$\gamma_\rho$ \end{tabular}}}}%
    \put(0,0){\includegraphics[width=\unitlength,page=7]{param_space_custom_targets.pdf}}%
    \put(0.04330838,0.06643121){\color[rgb]{0,0,0}\makebox(0,0)[lt]{\lineheight{1.25}\smash{\begin{tabular}[t]{l}0.10\end{tabular}}}}%
    \put(0,0){\includegraphics[width=\unitlength,page=8]{param_space_custom_targets.pdf}}%
    \put(0.04330838,0.14499327){\color[rgb]{0,0,0}\makebox(0,0)[lt]{\lineheight{1.25}\smash{\begin{tabular}[t]{l}0.12\end{tabular}}}}%
    \put(0,0){\includegraphics[width=\unitlength,page=9]{param_space_custom_targets.pdf}}%
    \put(0.04330838,0.22355533){\color[rgb]{0,0,0}\makebox(0,0)[lt]{\lineheight{1.25}\smash{\begin{tabular}[t]{l}0.14\end{tabular}}}}%
    \put(0,0){\includegraphics[width=\unitlength,page=10]{param_space_custom_targets.pdf}}%
    \put(0.04330838,0.3021174){\color[rgb]{0,0,0}\makebox(0,0)[lt]{\lineheight{1.25}\smash{\begin{tabular}[t]{l}0.16\end{tabular}}}}%
    \put(0,0){\includegraphics[width=\unitlength,page=11]{param_space_custom_targets.pdf}}%
    \put(0.04330838,0.38067946){\color[rgb]{0,0,0}\makebox(0,0)[lt]{\lineheight{1.25}\smash{\begin{tabular}[t]{l}0.18\end{tabular}}}}%
    \put(0,0){\includegraphics[width=\unitlength,page=12]{param_space_custom_targets.pdf}}%
    \put(0.04330838,0.45924153){\color[rgb]{0,0,0}\makebox(0,0)[lt]{\lineheight{1.25}\smash{\begin{tabular}[t]{l}0.20\end{tabular}}}}%
    \put(0.0322678,0.24830403){\color[rgb]{0,0,0}\rotatebox{90}{\makebox(0,0)[lt]{\lineheight{1.25}\smash{\begin{tabular}[t]{l}$R_w$ \end{tabular}}}}}%
    \put(0,0){\includegraphics[width=\unitlength,page=13]{param_space_custom_targets.pdf}}%
    \put(0.14246577,0.47625){\color[rgb]{0,0,0}\makebox(0,0)[lt]{\lineheight{1.25}\smash{\begin{tabular}[t]{l}$\mathrm{mean}(\mathrm{SNR_{im}}) (\mathrm{dB})$\end{tabular}}}}%
    \put(0,0){\includegraphics[width=\unitlength,page=14]{param_space_custom_targets.pdf}}%
    \put(0.56546549,0.04768663){\color[rgb]{0,0,0}\makebox(0,0)[lt]{\lineheight{1.25}\smash{\begin{tabular}[t]{l}1.0\end{tabular}}}}%
    \put(0,0){\includegraphics[width=\unitlength,page=15]{param_space_custom_targets.pdf}}%
    \put(0.62526191,0.04768663){\color[rgb]{0,0,0}\makebox(0,0)[lt]{\lineheight{1.25}\smash{\begin{tabular}[t]{l}1.5\end{tabular}}}}%
    \put(0,0){\includegraphics[width=\unitlength,page=16]{param_space_custom_targets.pdf}}%
    \put(0.68505833,0.04768663){\color[rgb]{0,0,0}\makebox(0,0)[lt]{\lineheight{1.25}\smash{\begin{tabular}[t]{l}2.0\end{tabular}}}}%
    \put(0,0){\includegraphics[width=\unitlength,page=17]{param_space_custom_targets.pdf}}%
    \put(0.74485476,0.04768663){\color[rgb]{0,0,0}\makebox(0,0)[lt]{\lineheight{1.25}\smash{\begin{tabular}[t]{l}2.5\end{tabular}}}}%
    \put(0,0){\includegraphics[width=\unitlength,page=18]{param_space_custom_targets.pdf}}%
    \put(0.80465118,0.04768663){\color[rgb]{0,0,0}\makebox(0,0)[lt]{\lineheight{1.25}\smash{\begin{tabular}[t]{l}3.0\end{tabular}}}}%
    \put(0,0){\includegraphics[width=\unitlength,page=19]{param_space_custom_targets.pdf}}%
    \put(0.8644476,0.04768663){\color[rgb]{0,0,0}\makebox(0,0)[lt]{\lineheight{1.25}\smash{\begin{tabular}[t]{l}3.5\end{tabular}}}}%
    \put(0.70233823,0.02395074){\color[rgb]{0,0,0}\makebox(0,0)[lt]{\lineheight{1.25}\smash{\begin{tabular}[t]{l}$\gamma_\rho$ \end{tabular}}}}%
    \put(0,0){\includegraphics[width=\unitlength,page=20]{param_space_custom_targets.pdf}}%
    \put(0.52846462,0.06643121){\color[rgb]{0,0,0}\makebox(0,0)[lt]{\lineheight{1.25}\smash{\begin{tabular}[t]{l}0.10\end{tabular}}}}%
    \put(0,0){\includegraphics[width=\unitlength,page=21]{param_space_custom_targets.pdf}}%
    \put(0.52846462,0.14499327){\color[rgb]{0,0,0}\makebox(0,0)[lt]{\lineheight{1.25}\smash{\begin{tabular}[t]{l}0.12\end{tabular}}}}%
    \put(0,0){\includegraphics[width=\unitlength,page=22]{param_space_custom_targets.pdf}}%
    \put(0.52846462,0.22355533){\color[rgb]{0,0,0}\makebox(0,0)[lt]{\lineheight{1.25}\smash{\begin{tabular}[t]{l}0.14\end{tabular}}}}%
    \put(0,0){\includegraphics[width=\unitlength,page=23]{param_space_custom_targets.pdf}}%
    \put(0.52846462,0.3021174){\color[rgb]{0,0,0}\makebox(0,0)[lt]{\lineheight{1.25}\smash{\begin{tabular}[t]{l}0.16\end{tabular}}}}%
    \put(0,0){\includegraphics[width=\unitlength,page=24]{param_space_custom_targets.pdf}}%
    \put(0.52846462,0.38067946){\color[rgb]{0,0,0}\makebox(0,0)[lt]{\lineheight{1.25}\smash{\begin{tabular}[t]{l}0.18\end{tabular}}}}%
    \put(0,0){\includegraphics[width=\unitlength,page=25]{param_space_custom_targets.pdf}}%
    \put(0.52846462,0.45924153){\color[rgb]{0,0,0}\makebox(0,0)[lt]{\lineheight{1.25}\smash{\begin{tabular}[t]{l}0.20\end{tabular}}}}%
    \put(0.51742405,0.24830404){\color[rgb]{0,0,0}\rotatebox{90}{\makebox(0,0)[lt]{\lineheight{1.25}\smash{\begin{tabular}[t]{l}$R_w$ \end{tabular}}}}}%
    \put(0,0){\includegraphics[width=\unitlength,page=26]{param_space_custom_targets.pdf}}%
    \put(0.66044472,0.47625){\color[rgb]{0,0,0}\makebox(0,0)[lt]{\lineheight{1.25}\smash{\begin{tabular}[t]{l}$\mathrm{mean} (r_{\Delta\boldsymbol{\sigma}})$\end{tabular}}}}%
    \put(0,0){\includegraphics[width=\unitlength,page=27]{param_space_custom_targets.pdf}}%
    \put(0.44357852,0.06643121){\color[rgb]{0,0,0}\makebox(0,0)[lt]{\lineheight{1.25}\smash{\begin{tabular}[t]{l}7.92\end{tabular}}}}%
    \put(0,0){\includegraphics[width=\unitlength,page=28]{param_space_custom_targets.pdf}}%
    \put(0.44357852,0.10777966){\color[rgb]{0,0,0}\makebox(0,0)[lt]{\lineheight{1.25}\smash{\begin{tabular}[t]{l}8.08\end{tabular}}}}%
    \put(0,0){\includegraphics[width=\unitlength,page=29]{param_space_custom_targets.pdf}}%
    \put(0.44357852,0.14912812){\color[rgb]{0,0,0}\makebox(0,0)[lt]{\lineheight{1.25}\smash{\begin{tabular}[t]{l}8.24\end{tabular}}}}%
    \put(0,0){\includegraphics[width=\unitlength,page=30]{param_space_custom_targets.pdf}}%
    \put(0.44357852,0.19047658){\color[rgb]{0,0,0}\makebox(0,0)[lt]{\lineheight{1.25}\smash{\begin{tabular}[t]{l}8.40\end{tabular}}}}%
    \put(0,0){\includegraphics[width=\unitlength,page=31]{param_space_custom_targets.pdf}}%
    \put(0.44357852,0.23182503){\color[rgb]{0,0,0}\makebox(0,0)[lt]{\lineheight{1.25}\smash{\begin{tabular}[t]{l}8.56\end{tabular}}}}%
    \put(0,0){\includegraphics[width=\unitlength,page=32]{param_space_custom_targets.pdf}}%
    \put(0.44357852,0.27317349){\color[rgb]{0,0,0}\makebox(0,0)[lt]{\lineheight{1.25}\smash{\begin{tabular}[t]{l}8.72\end{tabular}}}}%
    \put(0,0){\includegraphics[width=\unitlength,page=33]{param_space_custom_targets.pdf}}%
    \put(0.44357852,0.31452194){\color[rgb]{0,0,0}\makebox(0,0)[lt]{\lineheight{1.25}\smash{\begin{tabular}[t]{l}8.88\end{tabular}}}}%
    \put(0,0){\includegraphics[width=\unitlength,page=34]{param_space_custom_targets.pdf}}%
    \put(0.44357852,0.3558704){\color[rgb]{0,0,0}\makebox(0,0)[lt]{\lineheight{1.25}\smash{\begin{tabular}[t]{l}9.04\end{tabular}}}}%
    \put(0,0){\includegraphics[width=\unitlength,page=35]{param_space_custom_targets.pdf}}%
    \put(0.44357852,0.39721885){\color[rgb]{0,0,0}\makebox(0,0)[lt]{\lineheight{1.25}\smash{\begin{tabular}[t]{l}9.20\end{tabular}}}}%
    \put(0,0){\includegraphics[width=\unitlength,page=36]{param_space_custom_targets.pdf}}%
    \put(0.44357852,0.43856731){\color[rgb]{0,0,0}\makebox(0,0)[lt]{\lineheight{1.25}\smash{\begin{tabular}[t]{l}9.36\end{tabular}}}}%
    \put(0,0){\includegraphics[width=\unitlength,page=37]{param_space_custom_targets.pdf}}%
    \put(0.92873477,0.06643121){\color[rgb]{0,0,0}\makebox(0,0)[lt]{\lineheight{1.25}\smash{\begin{tabular}[t]{l}0.792\end{tabular}}}}%
    \put(0,0){\includegraphics[width=\unitlength,page=38]{param_space_custom_targets.pdf}}%
    \put(0.92873477,0.1100768){\color[rgb]{0,0,0}\makebox(0,0)[lt]{\lineheight{1.25}\smash{\begin{tabular}[t]{l}0.800\end{tabular}}}}%
    \put(0,0){\includegraphics[width=\unitlength,page=39]{param_space_custom_targets.pdf}}%
    \put(0.92873477,0.15372239){\color[rgb]{0,0,0}\makebox(0,0)[lt]{\lineheight{1.25}\smash{\begin{tabular}[t]{l}0.808\end{tabular}}}}%
    \put(0,0){\includegraphics[width=\unitlength,page=40]{param_space_custom_targets.pdf}}%
    \put(0.92873477,0.19736799){\color[rgb]{0,0,0}\makebox(0,0)[lt]{\lineheight{1.25}\smash{\begin{tabular}[t]{l}0.816\end{tabular}}}}%
    \put(0,0){\includegraphics[width=\unitlength,page=41]{param_space_custom_targets.pdf}}%
    \put(0.92873477,0.24101358){\color[rgb]{0,0,0}\makebox(0,0)[lt]{\lineheight{1.25}\smash{\begin{tabular}[t]{l}0.824\end{tabular}}}}%
    \put(0,0){\includegraphics[width=\unitlength,page=42]{param_space_custom_targets.pdf}}%
    \put(0.92873477,0.28465917){\color[rgb]{0,0,0}\makebox(0,0)[lt]{\lineheight{1.25}\smash{\begin{tabular}[t]{l}0.832\end{tabular}}}}%
    \put(0,0){\includegraphics[width=\unitlength,page=43]{param_space_custom_targets.pdf}}%
    \put(0.92873477,0.32830476){\color[rgb]{0,0,0}\makebox(0,0)[lt]{\lineheight{1.25}\smash{\begin{tabular}[t]{l}0.840\end{tabular}}}}%
    \put(0,0){\includegraphics[width=\unitlength,page=44]{param_space_custom_targets.pdf}}%
    \put(0.92873477,0.37195035){\color[rgb]{0,0,0}\makebox(0,0)[lt]{\lineheight{1.25}\smash{\begin{tabular}[t]{l}0.848\end{tabular}}}}%
    \put(0,0){\includegraphics[width=\unitlength,page=45]{param_space_custom_targets.pdf}}%
    \put(0.92873477,0.41559594){\color[rgb]{0,0,0}\makebox(0,0)[lt]{\lineheight{1.25}\smash{\begin{tabular}[t]{l}0.856\end{tabular}}}}%
    \put(0,0){\includegraphics[width=\unitlength,page=46]{param_space_custom_targets.pdf}}%
    \put(0.92873477,0.45924153){\color[rgb]{0,0,0}\makebox(0,0)[lt]{\lineheight{1.25}\smash{\begin{tabular}[t]{l}0.864\end{tabular}}}}%
    \put(0,0){\includegraphics[width=\unitlength,page=47]{param_space_custom_targets.pdf}}%
  \end{picture}%
\endgroup%

%% file: figures/E9_showcase.pdf_tex
\begingroup%
  \makeatletter%
  \providecommand\color[2][]{%
    \errmessage{(Inkscape) Color is used for the text in Inkscape, but the package 'color.sty' is not loaded}%
    \renewcommand\color[2][]{}%
  }%
  \providecommand\transparent[1]{%
    \errmessage{(Inkscape) Transparency is used (non-zero) for the text in Inkscape, but the package 'transparent.sty' is not loaded}%
    \renewcommand\transparent[1]{}%
  }%
  \providecommand\rotatebox[2]{#2}%
  \newcommand*\fsize{\dimexpr\f@size pt\relax}%
  \newcommand*\lineheight[1]{\fontsize{\fsize}{#1\fsize}\selectfont}%
  \ifx\svgwidth\undefined%
    \setlength{\unitlength}{889.3844845bp}%
    \ifx\svgscale\undefined%
      \relax%
    \else%
      \setlength{\unitlength}{\unitlength * \real{\svgscale}}%
    \fi%
  \else%
    \setlength{\unitlength}{\svgwidth}%
  \fi%
  \global\let\svgwidth\undefined%
  \global\let\svgscale\undefined%
  \makeatother%
  \begin{picture}(1,0.23104591)%
    \lineheight{1}%
    \setlength\tabcolsep{0pt}%
    \put(0,0){\includegraphics[width=\unitlength,page=1]{E9_showcase.pdf}}%
    \put(0.1109713,0.20217026){\color[rgb]{0,0,0}\makebox(0,0)[lt]{\begin{minipage}{0.52588791\unitlength}\raggedright \end{minipage}}}%
    \put(0.333621,0.21942198){\color[rgb]{0,0,0}\makebox(0,0)[lt]{\lineheight{1.25}\smash{\begin{tabular}[t]{l}$\boldsymbol{\mathrm{R}}_A$\end{tabular}}}}%
    \put(0.59539752,0.21942198){\color[rgb]{0,0,0}\makebox(0,0)[lt]{\lineheight{1.25}\smash{\begin{tabular}[t]{l}$\boldsymbol{\mathrm{R}}_B$\end{tabular}}}}%
    \put(0.85312107,0.21942198){\color[rgb]{0,0,0}\makebox(0,0)[lt]{\lineheight{1.25}\smash{\begin{tabular}[t]{l}$\boldsymbol{\mathrm{R}}_C$\end{tabular}}}}%
    \put(0.04318713,0.21942198){\color[rgb]{0,0,0}\makebox(0,0)[lt]{\lineheight{1.25}\smash{\begin{tabular}[t]{l}CT Image\end{tabular}}}}%
    \put(0.68408031,0.24696531){\color[rgb]{0,0,0}\makebox(0,0)[lt]{\begin{minipage}{2.7334043\unitlength}\raggedright \end{minipage}}}%
  \end{picture}%
\endgroup%